\documentclass[a4paper,11pt]{article}
\pdfoutput=1

\usepackage{jheppub, bm, color} 
\usepackage{amssymb,amsfonts,slashed,amsthm,amsmath,graphicx, soul}
\usepackage[caption=false]{subfig}
\usepackage{ulem}
\usepackage{hyperref}

\usepackage{epsfig}

\usepackage[compat=1.0.0]{tikz-feynman}
\usepackage{tikz-feynhand}

\newcommand{\lmk}{\left(}  
\newcommand{\rmk}{\right)}
\newcommand{\lkk}{\left[}  
\newcommand{\rkk}{\right]}

\newcommand{\del}{\partial}

\newcommand{\bea}{\begin{align}}
\newcommand{\eea}{\end{align}}
\newcommand{\beq}{\begin{eqnarray}}
\newcommand{\eeq}{\end{eqnarray}}

\newcommand{\dd}{\mathrm{d}}
\newcommand{\Mpl}{M_{\rm Pl}}

\newcommand{\cphi}{\varphi}

\newcommand{\eq}[1]{Eq.~(\ref{#1})}

\newcommand{\ms}{m}

\begin{document}

\begin{flushright}
TU-1197\\
KEK-QUP-2023-0013
\end{flushright}

\title{
Quantum decay of scalar and vector boson stars and oscillons into gravitons
}

\author{
Kazunori Nakayama$^{1,2}$, 
Fuminobu Takahashi$^{1}$, 
Masaki~Yamada$^{1,3}$
\\*[20pt]
$^1${\it \normalsize
Department of Physics, Tohoku University, Sendai, Miyagi 980-8578, Japan}  \\*[5pt]
$^2${\it \normalsize 
International Center for Quantum-field Measurement Systems for Studies of
the Universe and Particles (QUP, WPI), KEK, 1-1 Oho, Tsukuba, Ibaraki 305-0801, Japan}  \\*[5pt]
$^3${\it \normalsize 
FRIS, Tohoku University, Sendai, Miyagi 980-8578, Japan} %\\*[5pt]
}
\emailAdd{kazunori.nakayama.d3@tohoku.ac.jp}
\emailAdd{fumi@tohoku.ac.jp}
\emailAdd{m.yamada@tohoku.ac.jp}

\abstract{
We point out that a soliton such as an oscillon or boson star inevitably decays into gravitons through gravitational interactions. 
These decay processes exist even if there are no apparent self-interactions of the constituent field, scalar or vector, since they are induced by gravitational interactions.  Hence, our results provide a strict upper limit on the lifetime of oscillons and boson stars including the dilute axion star. 
We also calculate the spectrum of the graviton background from decay of solitons. 
}

\maketitle

%%%%%%%%%%%%%%%%%%%%%%%%%
\section{Introduction}
%%%%%%%%%%%%%%%%%%%%%%%%%

Bosons are ubiquitous in physics beyond the Standard Model. 
For example, the Peccei-Quinn (PQ) mechanism predicts a pseudo-Nambu-Goldstone boson, the QCD axion~\cite{Peccei:1977hh,Peccei:1977ur}. 
There may also exist a large number of axion-like particles in 
the low-energy effective field theory of the string theory. 
We generally call them axions for both cases. 
Even if their interaction rates with other particles are  suppressed by a large decay constant, 
they can be produced via coherent oscillations in the early Universe~\cite{Preskill:1982cy,Abbott:1982af,Dine:1982ah} 
and are a good candidate for dark matter (DM). 
On another hand, massive vector fields in the dark sector can also be considered as a candidate for DM (see Ref.~\cite{Antypas:2022asj} and references therein). However, the production mechanism of vector DM is non-trivial, and includes the misalignment mechanism~\cite{Nelson:2011sf,Arias:2012az} (see, however, Ref.~\cite{Nakayama:2019rhg, Nakayama:2020rka} for its difficulty and Ref.~\cite{Kitajima:2023fun} for a viable scenario), emission from cosmic strings~\cite{Long:2019lwl,Kitajima:2022lre}, 
phase transition~\cite{Nakayama:2021avl}, 
scalar-field dependent gauge kinetic coupling~\cite{Salehian:2020asa,Firouzjahi:2020whk}, 
preheating and decay of axion-like particles~\cite{Agrawal:2018vin,Dror:2018pdh,Co:2018lka,Bastero-Gil:2018uel,Co:2021rhi,Kitajima:2023pby}, 
and gravitational production during and after inflation~\cite{Graham:2015rva,Ema:2019yrd,Ahmed:2020fhc,Nakayama:2020ikz,Kolb:2020fwh,Arvanitaki:2021qlj,Sato:2022jya}. 
The cosmological implications of those bosons are therefore important to understand the consistent scenario of particle physics throughout the history of the Universe.

In cosmology, light bosons are usually produced with extremely large occupancy numbers, which imply the formation of Bose-Einstein condensate~\cite{Guth:2014hsa}. In fact, gravitationally bound clumps, called boson stars or oscillatons~\cite{Kaup:1968zz, Ruffini:1969qy, Colpi:1986ye, Seidel:1991zh, Tkachev:1991ka, Kolb:1993zz, Kolb:1993hw}, are formed via the growth of perturbations during the matter dominated era~\cite{Schive:2014hza,Levkov:2018kau,Widdicombe:2018oeo,Eggemeier:2019jsu,Chen:2020cef}. 
In the context of axion cosmology, gravitationally bounded clumps of axion are sometimes called dilute axion stars or axiton~\cite{Kolb:1993hw, Braaten:2015eeu, Braaten:2016dlp, Visinelli:2017ooc,Schiappacasse:2017ham}. 
In this paper, we simply call them as boson stars.
The formation of such objects is supported by detailed numerical simulations~\cite{Veltmaat:2018dfz,Eggemeier:2019jsu,Chen:2020cef}, 
where it was found that localized objects of scalar fields are formed at the center of DM halo, for the QCD axion and ultralight (string) axion.
The configuration and properties of the boson star can be calculated by the approach used in Refs.~\cite{Chavanis:2011zi,Chavanis:2011zm} (see also Ref.~\cite{Eby:2018ufi}). 
The boson star is not completely stable but has a long lifetime in vacuum on cosmological time scales~\cite{Mukaida:2016hwd, Braaten:2016dlp,Visinelli:2017ooc, Eby:2018ufi}. 
The evaporation and growth of boson stars in the DM halo was investigated in Ref.~\cite{Chan:2022bkz}. 
Those localized objects may affect the image of gravitational lensing~\cite{Fujikura:2021omw,Ellis:2022grh}. 
Those objects are therefore relevant in cosmology, especially when they are dominant components of DM in the Universe.

On the other hand, if the attractive self interaction of the scalar field is stronger than the gravitational interaction, 
the localized object is called an oscillon or I-ball (or dense axion star when the scalar is axion)~\cite{Bogolyubsky:1976nx,Segur:1987mg,Gleiser:1993pt,Copeland:1995fq,Gleiser:1999tj,Honda:2001xg,Kasuya:2002zs,Fodor:2006zs,Fodor:2008du,Gleiser:2008ty,Gleiser:2009ys,Amin:2010jq,Amin:2011hj,Salmi:2012ta,Saffin:2014yka,Mukaida:2014oza,Kawasaki:2015vga}. 
A perturbative expansion in terms of the spatial gradient or binding energy is often used to determine the configuration of oscillons~\cite{Fodor:2008es}, whereas some systematic approaches to calculate the configuration in the non-relativistic effective field theory are also proposed in Refs.~\cite{Braaten:2016kzc,Mukaida:2016hwd,Namjoo:2017nia,Braaten:2018lmj,Levkov:2022egq}. 
Those objects are realized by the QCD axion after the QCD phase transition in the post-inflationary PQ symmetry breaking scenario~\cite{Vaquero:2018tib} and in a large misalignment scenario~\cite{Arvanitaki:2019rax}. 
The inflaton may also form oscillons after inflation, depending on models (see, e.g., Ref.~\cite{Copeland:2002ku,Amin:2011hj}). 
However, the lifetime of oscillons tends to be much shorter than the cosmological time scale. 
It was noted that the amplitude of the primordial gravitational waves may be enhanced 
if the oscillons dominate the Universe before they decay~\cite{Inomata:2019ivs,Lozanov:2022yoy}. 
Therefore, determining the lifetime of those localized objects has important implications for cosmological observations. 

Recently, the formation of similar extended objects from vector bosons, which we call the vector boson star or the vector oscillon 
depending on the attractive force that stabilizes the configuration, has been discussed~\cite{Adshead:2021kvl,Jain:2021pnk,Zhang:2021xxa,Amin:2023imi}.
Detailed numerical simulations~\cite{Gorghetto:2022sue,Amin:2022pzv,Jain:2023ojg} supports the formation of vector boson stars. 
They may also have lots of phenomenological impacts, and it is important to understand the stability of these objects.

The classical/quantum stability of the boson star or the oscillon is extensively studied in the literature.
In Refs.~\cite{Mukaida:2016hwd,Eby:2018ufi}, a diagrammatic technique to calculate the {\it classical} decay rate of oscillons and axion stars was developed.  
The resulting decay rate is consistent with that of numerical simulations of oscillons~\cite{Salmi:2012ta,Mukaida:2016hwd}. 
See also Refs.~\cite{Eby:2015hyx,Eby:2017azn,Ibe:2019vyo,Ibe:2019lzv,Zhang:2020bec,Zhang:2020ntm} for the analysis of the oscillon decay, which agrees with Refs.~\cite{Mukaida:2016hwd,Eby:2018ufi} and a similar formula was derived there from different perspectives. 
In Refs.~\cite{Grandclement:2011wz,Fodor:2009kg,Page:2003rd}, they considered the classical decay rate of oscillatons. 
In particular, the dilute axion star is expected to have a very long lifetime. 
In Ref.~\cite{Zhang:2020ntm, Eby:2020ply}, they discussed the effect of gravity on the classical decay rate of axion stars.

On the other hand, the {\it quantum} decay of oscillons and axion stars was considered in Refs.~\cite{Hertzberg:2010yz,Kawasaki:2013awa,Braaten:2016dlp,Hertzberg:2018zte} with and without parametric resonance. 
Those works particularly considered the decay into generic scalar fields or photons. 
The quantum decay process corresponds to the imaginary part of the loop diagrams in the above diagrammatic technique.

In this paper, we consider the {\it quantum} decay of soitons, which collectively represent
scalar/vector boson stars and oscillons,
into gravitons via gravitational interactions. 
Although a soliton is spherically symmetric, it can produce gravitons (or gravitational waves) quantum mechanically. 
This is similar to the graviton emission from the spherically symmetric black holes via Hawking radiation, which is a semi-classical process and is classically forbidden.
The quantum decay into graviton is inevitable even if the boson has only the quadratic potential. 
A free real scalar field is the simplest model of quantum field theory but is non-trivial due to the gravitational interactions, where it is localized by its self gravitational potential and decays into gravitons.

Gravitational particle production due to an oscillating scalar field in a cosmological setup has been discussed in Ref.~\cite{Ema:2015dka}, which is also interpreted as the gravitational annihilation of the scalar field into any field. Later this and related subjects have been studied in many papers in the context of dark matter production and reheating~\cite{Ema:2015dka,Ema:2016hlw,Tang:2017hvq,Ema:2018ucl,Chung:2018ayg,Ema:2019yrd,Mambrini:2021zpp,Basso:2021whd,Clery:2021bwz,Haque:2021mab}, graviton production~\cite{Ema:2015dka,Schiappacasse:2016nei,Ema:2016hlw,Ema:2020ggo}, gravitational annihilation of Standard Model particles in thermal bath~\cite{Garny:2015sjg,Tang:2016vch,Tang:2017hvq,Garny:2017kha,Gross:2020zam}.\footnote{See also Refs.~\cite{Ford:1986sy,Chung:2001cb,Chung:2011ck,Graham:2015rva,Hashiba:2018tbu,Ema:2018ucl,Ema:2019yrd,Ahmed:2020fhc,Kolb:2020fwh} for gravitational production during inflation or transition from the inflationary epoch.}
With this knowledge, it is not surprising that the oscillating scalar field inside the soliton leads to the production of any particle through the gravitational process like $\phi\phi\to XX$, where $\phi$ is a scalar or vector that constitutes the soliton and $X$ denotes any light field since they are inevitably coupled to $\phi$ through gravity.
In practice, production of fermions and massless vector bosons are suppressed and dominant process is that into the graviton pair or the Higgs boson (or longitudinal gauge boson) pair if $\phi$ is heavier than them.
In particular, the graviton production process is always present no matter how $\phi$ is light. Thus this process provides a strict upper limit on the lifetime of solitons. 
We stress that this process is inevitable even if there is no explicit scalar self-interaction term in the action.
We estimate the graviton production rate by deriving the graviton equation of motion under the scalar/vector soliton background and finding that it is the same as the Mathieu equation. The source of the oscillating term is identified with the oscillating gravitational potential in the presence of soliton.

To the best of our knowledge, the only previous work that explicitly pointed out the quantum graviton production from the scalar oscillon or boson star is Ref.~\cite{Page:2003rd}. There the graviton production rate is calculated by interpreting it as the annihilation process $\phi\phi\to hh$, where $h$ denotes the graviton. In this paper we estimate the graviton production rate by using the time-dependent gravitational potential with the classical soliton configuration.  
Our approach is useful for more general objects such as Q-balls~\cite{Coleman:1985ki}, vector boson stars or vector oscillons, since in these multi-field solitons it is sometimes nontrivial whether such a perturbative annihilation picture is applicable or not~\cite{Page:2003rd}. 
In particular, we will see an example in which even the graviton production is absent and our formulation makes it clear.

This paper is organized as follows. In Sec.~\ref{sec:boson} we briefly review properties of scalar solitons, i.e.  oscillons/boson stars for a scalar field, and estimate the time-dependence of the gravitational potential.
In Sec.~\ref{sec:vector} we discuss the case of vector solitons.
In Sec.~\ref{sec:decay} we estimate the gravitational {\it quantum} decay rate of a scalar/vector soliton, by using the information on the gravitational potential derived in preceding sections.
In Sec.~\ref{sec:application} we point out several phenomenological impacts.
Sec.~\ref{sec:conclusion} is devoted to conclusions and discussion.

%%%%%%%%%%%%%%%%%%%%%%%%%%%%
\section{Scalar solitons}
\label{sec:boson}
%%%%%%%%%%%%%%%%%%%%%%%%%%%%

We mainly focus on a spherically symmetric localized configuration of a free real scalar or vector field with gravity, which is sometimes called a boson star. In the context of axion, this corresponds to a dilute axion star. 
Our qualitative result for quantum decay rate can be applied to almost any localized scalar-field configurations, including oscillons and axion stars. For concreteness, we mainly consider boson stars in this paper 
and briefly comment on the case of oscillons. 
We consider a scalar field in this section and a vector field in Sec.~\ref{sec:vector}. 

%%%%%%%%%%%%%%%%%%%%%%%%%%%%
\subsection{Setup}
\label{sec:setup}
%%%%%%%%%%%%%%%%%%%%%%%%%%%%

We consider a system with the action of 
\beq
 S = \int \sqrt{-g} \, d^4 x \lkk -\frac{1}{16\pi G} R + \frac12 g^{\mu \nu} \del_\mu \phi \del_\nu \phi - V(\phi) \right]\,,
\eeq
where $R$ is the Ricci scalar, $g$ is the determinant of the metric, 
$G$ ($=(8\pi \Mpl^2)^{-1}$) is the gravitational constant, 
and $M_{\rm Pl}$ is the reduced Planck mass. 
The scalar potential is given by 
\beq
 V(\phi) 
 &=& 
 \frac12 \ms^2 \phi^2 - \frac{1}{24}\lambda \phi^4 
 + \dots\,, 
\eeq
where $\ms$ is the scalar mass and $\lambda$ is a quartic coupling. 
We consider the case with a sufficiently small $\lambda$. 
For the case of axion with a sine-Gordon potential, 
one can identify $\lambda = \ms^2 / f_a^2$ with $f_a$ being the axion decay constant, neglecting higher-dimensional terms.

The gravitational interaction is a long-range force, whereas 
the self-interaction is a point interaction. 
The former one can be dominant to determine the localized configuration, if the density at the center of the boson star is sufficiently low. 
The extent to which this approximation is valid can be understood by estimating the gradient energy, the gravitational potential energy, and the potential energy of the self-interaction,
\beq
 &&\delta_x \sim \frac{\lmk \nabla \phi \rmk^2}{\ms^2 \phi^2} \sim \frac{1}{(\ms R)^2}\,, 
 \label{eq:deltax}
 \\
 &&\delta_g \sim \frac{1}{\Mpl^2} \int d^3 x \frac{\ms^2 \phi^2}{r} \sim \frac{\ms^2 \phi^2}{\Mpl^2} R^2\,, 
 \label{eq:deltag}
 \\
 &&\delta_\phi \sim \frac{\lambda \phi^4}{ \ms^2 \phi^2}\,, 
\eeq
where we have normalized the energy densities by $\ms^2 \phi^2$, 
 $R$ represents the typical size of the boson star and $\phi$ represents the boson field value at the center of boson star.\footnote{
This is an abuse of notation, but this $R$ should not be confused with the Ricci scalar; the Ricci scalar only appears in the Hilbert-Einstein action in this paper.
 } 
The self-interaction is negligible if $\delta_g \gg \delta_\phi$. 
In this case, the size of boson star is determined by $\delta_x \sim \delta_g$ 
and we obtain $R \sim (\Mpl / \phi)^{1/2}/\ms$. 
Using this relation, the consistency condition for this approximation, $\delta_g \gg \delta_\phi$, is written as
\beq
 \lambda \ll \frac{\ms^4 R^2}{\Mpl^2} \sim \frac{\ms^2 }{ \phi \Mpl}\,. 
\eeq
For the case of axion, 
this condition implies
\beq
 f_a \gg \sqrt{\phi \Mpl}\,. 
 \label{app}
\eeq
We assume that those conditions are met when we discuss boson stars.

Since the gravitational interaction is important, 
we write the metric around the boson star such as
\begin{equation}
 \dd s^2 = e^{-2\Psi(t,r)}\,\dd t^2 - e^{2 \Phi(t,r)} \lkk \dd r^2 + r^2\left(\dd\theta^2 + \sin^2\theta\,\dd\varphi^2\right) \rkk\,. 
\end{equation}
Here, we adopt the isotropic coordinates, which should not be confused with the spherical coordinates~\cite{Friedberg:1986tp}.\footnote{
	Note that the spherical coordinates has been used in Refs.~\cite{Eby:2018ufi,Zhang:2020ntm}. The reason why we are using the isotropic coordinates is that it is directly compared with the Cartesian coordinates, which will be used in the analysis of particle production in Sec.~\ref{sec:decay}.
}
The energy-momentum tensor is written as 
\beq
 T_{\mu \nu} &&= \del_\mu \phi \del_\nu \phi - \frac12 g_{\mu \nu} g^{\rho \sigma} \del_\rho \phi \del_\sigma \phi
 + g_{\mu\nu} V (\phi)\,, 
 \label{EMtensor}
\eeq
where 
$V(\phi) \simeq \frac{1}{2}\ms^2 \phi^2$.

%%%%%%%%%%%%%%%%%%%%%%%%%%%%
\subsection{Configuration of scalar boson star}
%%%%%%%%%%%%%%%%%%%%%%%%%%%%

The Einstein equations read
\begin{align}
	&\displaystyle {G^0}_0=
	-e^{-2 \Phi} 
	\left[ 2 \frac{\partial^2\Phi}{\partial r^2} + \frac{1}{r}
	\lmk 4+r \frac{\partial\Phi}{\partial r} \rmk \frac{\partial\Phi}{\partial r} 
\right] + 3 e^{2 \Psi} \dot{\Phi}^2
	=\frac{\rho}{M_{\rm Pl}^2}, \label{G00}\\
	&\displaystyle {G^0}_r=
	-2 e^{2\Psi} \lmk \frac{\partial\Psi}{\partial r} \dot{\Phi} + \frac{\partial \dot{\Phi}}{\partial r} \rmk
	= \frac{e^{2\Psi}}{M_{\rm Pl}^2}\frac{\partial\phi}{\partial t}\frac{\partial\phi}{\partial r}, \label{G01}\\
	&\displaystyle {G^r}_r=
	e^{-2 \Phi} \left[
	\frac{2}{r} \frac{\partial\Psi}{\partial r} - \frac{2}{r} \frac{\partial\Phi}{\partial r} 
	+ 2 \frac{\partial\Psi}{\partial r} \frac{\partial\Phi}{\partial r} 
	- \lmk \frac{\partial\Phi}{\partial r} \rmk^2  
	\right]
	+ e^{2 \Psi } \lmk 3 \dot{\Phi}^2 + 2 \dot{\Phi}\dot{\Psi} 
	+ 2 \ddot{\Phi} \rmk 	
    =-\frac{p_r}{M_{\rm Pl}^2}\,.
    \label{G11}
\end{align}
The energy density and pressure are defined as
\begin{align}
	&\displaystyle\rho\equiv {T^{0}}_0 = \frac{e^{2\Psi}}{2}\dot{\phi}^2+  \frac{e^{-2\Phi}}{2}\left(\frac{\partial\phi}{\partial r}\right)^2 + \frac{1}{2} \ms^2\phi^2\,,\\
	&\displaystyle p_r \equiv -{T^r}_r = \frac{e^{2\Psi}}{2} \dot{\phi}^2+  \frac{e^{-2\Phi}}{2}\left(\frac{\partial\phi}{\partial r}\right)^2 -\frac{1}{2} \ms^2\phi^2\,,
\end{align}
where we have used $V(\phi) \simeq \frac{1}{2} m^2 \phi^2.$

We are interested in the weak gravity limit, where we can treat the gravity as perturbations. 
We thus write $e^{-2\Psi} \simeq 1 - 2\Psi(t,r)$ and $e^{2\Phi} \simeq 1 + 2 \Phi(t,r)$ and treat $\Psi$ and $\Phi$ as small parameters. From the Einstein equation, they obey 
\beq
 &&\Psi(t,r) =  G \int_r^\infty \frac{\dd r'}{r'^2} \lmk M(r') + 4 \pi r'^3  p_r(t,r')  \rmk + \int_r^\infty \dd r' \, r' \ddot{\Phi}\,, 
 \label{Psi}
 \\
 &&\Phi(t,r) = G \int_r^\infty \frac{\dd r'}{r'^2} M(r')\,,
 \label{Phi}
\eeq
where 
\beq
 M(r) \equiv \int_0^r dr' \, 4 \pi r'^2 \rho(r')\,, 
\eeq
is the energy enclosed within the radius $r$. 
As we shall see later, the time dependence of
$M(r)$ is negligible for our purpose.
Also, while it slowly changes with time via the decay of the scalar field as we will see in Sec.~\ref{sec:application}, 
its time scale is much longer. 
Thus we can neglect $\dot{\Phi}$ to derive the boson star configuration.

The equation of motion for a free scalar field $\phi$ is given by
\begin{align}
	e^{2\Psi}\left[ \ddot\phi +(\dot\Psi+3\dot\Phi)\dot\phi\right] -e^{-2\Phi}\left[ \frac{\partial^2\phi}{\partial r^2} +\frac{2}{r}\frac{\partial\phi}{\partial r}-	
	\left(\frac{\partial \Psi}{\partial r}- \frac{\partial \Phi}{\partial r}\right)\frac{\partial\phi}{\partial r}\right] + 
	\frac{\partial V}{\partial \phi}=0\,. 
\end{align}
The scalar field can have a localized configuration thanks to the gravitational attractive interaction. This implies that the spatial gradient energy of the scalar field is proportional to the gravitational potential. 
We thus neglect the cross term between the spatial gradient and the gravitational potential 
in the weak gravity limit. 
Moreover, as we will see shortly, the time-dependent parts of $\Psi$ and $\Phi$ are next-to-leading terms and we can neglect their time derivatives in the equation of motion.
We then obtain the simplified equation of motion,
\begin{align}
	(1 + 2 \Psi_0) \ddot\phi 
	- \lmk \frac{\partial^2\phi}{\partial r^2} +\frac{2}{r}\frac{\partial\phi}{\partial r} \rmk 
    + \ms^2 \phi = 0\,, 
\label{eq:EOM}
\end{align}
where we denote the leading-order (time-independent) part of $\Psi$ as $\Psi_0$.

Now, let us take an ansatz $\phi(t,r) = \phi_r(r) \cos (\omega t)$ to solve the equation of motion. 
The equation of motion for $\phi_r(r)$ is then given by 
\begin{align}
	\frac{\del^2\phi_r}{\del r^2} + \frac{2}{r}\frac{\del\phi_r}{\del r}
         		- \lkk m^2 - \omega^2 (1+2  \Psi_0(r)  ) \rkk \phi_r =0\,, 
         		\label{eq:phir}
\end{align}
where $\Psi_0(r)$ is determined from 
\beq
\frac{1}{4\pi G} \left(\frac{\del^2 \Psi_0 }{\del r^2} + \frac{2}{r}\frac{\del  \Psi_0  }{\del r}\right)= 
-\frac{1}{2} m^2 \phi_r^2(r)\,. 
\label{eq:Psi}
\eeq

To obtain the configuration $\phi_r (r)$ through numerical calculations, a convenient approach is to rescale the variables into dimensionless units.
The set of equations (\ref{eq:phir}) and (\ref{eq:Psi}) 
can be rewritten as 
\begin{align}
	&\frac{\del^2\cphi}{\del z^2} + \frac{2}{z}\frac{\del\cphi}{\del z}
         		+ \lkk 1 - g(z) \rkk \cphi =0, 
         		\label{eq:cphi}
\\
&\frac{\del^2  g }{\del z^2} + \frac{2}{z}\frac{\del g }{\del z} = \cphi^2(z), 
\label{eq:g}
\end{align}
where we define
\begin{align}
    &R \equiv \frac{1}{\sqrt{\omega^2 - \ms^2 + 2 \omega^2 \Psi_0 (0) }}\,,
    \\
    &r = R z\,, 
    \\
    &\phi_r (r)= \frac{\cphi(z)}{\sqrt{4 \pi G} \, \ms \omega R^2 } \,,
    \\
    &\Psi_0 (r) = \Psi_0 (0) - \frac{g(z)}{2 \omega^2 R^2}\,. 
\end{align}
Here we note that $ \Psi_0 = 0$ at an asymptotic infinity, which implies $g_\infty \equiv g(\infty) = 2 \omega^2 R^2 \Psi_0 (0)$. 
We numerically solve the above equations from $z = z_0 \ll 1$ to $z = z_1 \gg 1$ 
with the boundary condition of 
$\cphi (z_0) = \cphi_*$, $\cphi'(z_0) = -\cphi_* z_0 /2$, $g(z_0) = 0$, $g'(z_0) = \cphi_*^2 z_0 / 2$, and $\cphi(\infty) = 0$. 
The localized configuration is realized only by a certain value of $\cphi_*$, which can be determined by the shooting algorithm. 
The result is given by 
\begin{align}
 &\cphi_* \simeq 1.089, 
 \\
 &g(z \gg 1) \simeq 2.066 - \frac{3.619}{z}\,. 
 \label{eq:g-solution}
\end{align}
In particular, we obtain $g_\infty \simeq 2.066$, which implies 
\begin{align}
 &R \simeq \frac{\sqrt{g_\infty - 1}}{\epsilon \,  m} \simeq \frac{1.0 }{\epsilon \, m}\,, 
 \\
 &\phi_r (0) \simeq \frac{0.29 \, \epsilon^2}{\sqrt{G}}\,,
\end{align}
where we define $\epsilon^2 \equiv (m^2-\omega^2)/m^2$ ($\ll 1$).

We define $R_{99}$ by the radius at which 99\% of total energy of the axion star is enclosed. 
The total energy and $R_{99}$ are given by 
\begin{align}
 &E \equiv M(\infty) \simeq 1.8 \,  \frac{\epsilon}{G m} \,,
 \label{eq:E}
 \\
 &R_{99} \simeq 5.2 \, R \simeq \frac{5.4}{\epsilon \, m}\,. 
\end{align}

We note that 
the amplitude $\phi_r(0)$ has to be much larger than $\ms$, since otherwise the number density is too small for the classical picture to be valid. This requires $\epsilon \gg \sqrt{m/\Mpl}$.

%%%%%%%%%%%%%%%%%%%%%%%%%%%%
\subsection{Oscillating behavior of the gravitational potential}
%%%%%%%%%%%%%%%%%%%%%%%%%%%%

Here we derive the the behavior of gravitational potential that helps to understand the quantum decay process. 
The potential $\Phi$ is given in Eq.~(\ref{Phi}) as a solution to Eq.~(\ref{G00}). To find a time dependence of $\Phi$, it is convenient to look at Eq.~(\ref{G01}). By substituting an ansatz for the configuration $\phi(t,r)=\phi_r(r)\cos(\omega t)$, it is easy to find that the solution is given in the form of
\begin{align}
	\Phi(t,r) \simeq \Phi_0(r) - \frac{\phi_r(r)^2}{16M_{\rm Pl}^2}\cos(2\omega t)\,,  \label{Phi_osc}
\end{align}
where $\Phi_0(r)$ is time-independent part. Parametrically $|\Phi_0| \sim (R \omega \phi_r/M_{\rm Pl})^2$ and it is $(R \omega)^2$ times larger than the oscillating term for a typical oscillon radius $R$.\footnote{
	The expression (\ref{Phi_osc}) is independent of whether the object is stabilized by the self interaction (oscillon) or gravitational force (boson star). Thus we can use the oscillating gravitational potential of $\sim (\phi_r(r)^2/16M_{\rm Pl}^2)\cos(2\omega t)$ in both cases to estimate the graviton production rate. See Sec.~\ref{sec:decay}.
}
As a consistency check, the same expression can be directly derived from (\ref{Phi}). To see this, let us note that the energy density $\rho$ satisfies the equation
\begin{align}
	\dot \rho = e^{-2\Phi} \left[ \frac{1}{r^2}\frac{\partial}{\partial r}\left(r^2\dot\phi \frac{\partial \phi}{\partial r} \right) 
	+ \frac{\partial \phi}{\partial r} \left( \dot\phi \left(\frac{\partial \Phi}{\partial r} -\frac{\partial \Psi}{\partial r} \right) -\dot\Phi \frac{\partial \phi}{\partial r} \right) \right] - e^{2\Psi} 3\dot\Phi \dot\phi^2\,.
\end{align}
Neglecting terms involving $\Psi$ or $\Phi$, we find
\begin{align}
	\rho(t,r) \simeq \rho_0(r) + \frac{1}{8r^2}\frac{\partial}{\partial r}\left(r^2\frac{\partial \phi_r^2}{\partial r} \right) \cos(2\omega t)\,.
\end{align}
By substituting this into Eq.~(\ref{Phi}) we obtain the same expression as (\ref{Phi_osc}).
From Eq.~(\ref{G11}), we also obtain the time-dependence of $\Psi$ as
\begin{align}
	\Psi(t,r) \simeq \Psi_0(r) + \Psi_1(r)\cos(2\omega t)\,,  \label{Psi_osc}
\end{align}
where $|\Psi_0| \sim (R \omega \phi_r/M_{\rm Pl})^2$ and $|\Psi_1| \sim (\phi_r/M_{\rm Pl})^2$.
The time dependence of the gravitational potential is essential to understand the quantum decay of the boson star, i.e., the production of gravitationally coupled particles such as gravitons as we will see in Sec.~\ref{sec:decay}.

%%%%%%%%%%%%%%%%%%%%%%%%%%%%
\subsection{The case of complex scalar: Q-balls}  \label{sec:Q-ball}
%%%%%%%%%%%%%%%%%%%%%%%%%%%%

Here we give comments on the case of a complex scalar $\phi$ with global U(1) symmetry $\phi\to \phi e^{i\theta}$ with an arbitrary constant $\theta$. The action is
\beq
 S = \int \sqrt{-g} \, d^4 x \lkk -\frac{1}{16\pi G} R +g^{\mu \nu} \del_\mu \phi^* \del_\nu \phi - V(|\phi|) \right]\,.
\eeq
In this case, it is known that there appear soliton-like objects, called Q-balls~\cite{Coleman:1985ki,Cohen:1986ct}, if the potential is shallower than the quadratic potential. Note that, in contrast to oscillons/I-balls, the potential does not have to be dominated by a quadratic term~\cite{Kasuya:2002zs}. 
Q-balls are known to play essential roles in the dynamics of supersymmetric flat directions~\cite{Kusenko:1997ad,Kusenko:1997zq,Kusenko:1997si} in the context of Affleck-Dine baryogenesis~\cite{Affleck:1984fy,Dine:1995kz}.

The Q-ball solution is written as $\phi(t,r) = \phi_r(r)e^{i\omega t}$ with a real function $\phi_r(r)$. 
A crucial difference from the case of real scalar is that it allows a solution in which the gravitational potential $\Phi$ does not oscillate with time.
It is easily seen from the complex scalar version of (\ref{G01}):
\begin{align}
	\displaystyle {G^0}_r=
	-2 e^{2\Psi} \lmk \frac{\partial\Psi}{\partial r} \dot{\Phi} + \frac{\partial \dot{\Phi}}{\partial r} \rmk
	= \frac{e^{2\Psi}}{2M_{\rm Pl}^2}\left(\frac{\partial\phi^*}{\partial t}\frac{\partial\phi}{\partial r}
	+\frac{\partial\phi}{\partial t}\frac{\partial\phi^*}{\partial r}\right)\,.
	\label{G01_complex}
\end{align}
For the Q-ball solution, 
the right hand side is zero and hence the gravitational potential is constant with time. 
Therefore, no gravitational particle production happens in this limit and such a configuration is stable even against to the graviton production (see Sec.~\ref{sec:decay}).
It can be understood as a consequence of the U(1) symmetry and the conserved global charge associated with it.
If the orbit is initially elliptical in the complex plane, it will approach to the circular orbit by quantum decay process, including the graviton emission even if there are no explicit U(1) breaking terms. 

One may add Planck-suppressed higher-dimensional terms that explicitly break U(1) symmetry. They are motivated by a low-energy effective field theory for quantum gravity, where there should be no global symmetry in nature. 
Those U(1) symmetry breaking terms distort the dynamics of the complex scalar field from the circular orbit in the complex plane. Then the gravitational potential $\Phi$ has an oscillating term that can result in gravitational emission. 
As the amplitude of scalar field decreases, 
the effect of higher-dimensional U(1) breaking terms becomes weaker. Therefore the graviton emission rate becomes smaller as the Q-ball emits more gravitons.

%%%%%%%%%%%%%%%%%%%%%%%%%%%%%%%%%%%%
\section{Vector solitons} 
\label{sec:vector}
%%%%%%%%%%%%%%%%%%%%%%%%%%%%%%%%%%%%

In this section we consider vector solitons and discuss the properties of the gravitational potentials, particularly focusing on their time dependence.
While the most discussion is parallel to the case of scalar solitons discussed in the previous section, a crucial difference is that there are polarization degrees of freedom in the case of vector solitons. 
We follow the analysis in Ref.~\cite{Adshead:2021kvl} (see also Ref.~\cite{Jain:2021pnk}) and extend it to calculate the time-dependence of gravitational potential.

%%%%%%%%%%%%%%%%%%%%%
\subsection{Setup}
%%%%%%%%%%%%%%%%%%%%%

The action we consider is given by 
\beq
\label{eq:Action}
S=\int d^4x\sqrt{-g}\left[-\frac{R}{16\pi G}-\frac{1}{4}F^{\mu\nu}F_{\mu\nu}+\frac{1}{2}m^2A^{\mu}A_{\mu}\right]\,,
\eeq
where $m$ is the Stuckelberg mass for the dark photon $A^\mu$ 
and $F_{\mu \nu} = \del_\mu A_\nu - \del_\nu A_\mu$ is its field strength. 
The energy-momentum tensor is then given by 
\beq
\label{eq:EMtensor}
T^\mu{}_\nu=m^2A^{\mu} A_{\nu}-F^{\mu\alpha}F_{\nu\alpha} +\delta^\mu_\nu\left[-\frac{m^2}{2}A_\alpha A^\alpha+\frac{1}{4}F^{\alpha\beta}F_{\alpha\beta}\right]\,.
\eeq

The Maxwell equations read 
\beq
\label{eq:Maxwell}
\nabla_{\mu}F^{\mu\nu}
= \frac{1}{\sqrt{-g}} \del_\mu 
 \lmk \sqrt{-g} F^{\mu\nu} \rmk 
=-m^2A^\nu\,,
\eeq
where we note that 
\beq
 F^{\mu \nu} = g^{\mu \alpha} \lmk \del_\alpha A^\nu 
 - A_\beta \del_\alpha g^{\nu \beta} \rmk 
 - g^{\nu \alpha} \lmk \del_\alpha A^\mu 
 - A_\beta \del_\alpha g^{\mu \beta} \rmk \,.
\eeq
This particularly implies the Lorentz condition
\beq
\label{eq:Lorentz1}
\nabla_{\nu}A^{\nu}=0\,.
\eeq
Since we consider a massive gauge field, 
there are three degrees of freedom for the gauge field and 
$A^0$ is a non-dynamical field.

We are interested in the weak-gravity limit, where the metric is treated as a perturbation from the Minkowski metric. 
In the Newtonian gauge, we write 
\beq
ds^2=(1-2\Psi)d\tau^2+2V_i dx^id\tau
-\left[(1+2\Phi)\delta_{ij} -h_{ij}\right]dx^idx^j\,,
\eeq
where $\del_i V_i = 0$ and we also choose a gauge to eliminate the transverse vector degree of freedom. 
The transverse-traceless metric perturbation  $h_{ij}$ satisfies $\partial_i h_{ij} = \partial_i h_{ji} = 0$, and $h_{ii} = 0$. 
This completely fixes the gauge: $\Psi$, $\Phi$ and $V_i$ are given in terms of $A_\mu$.
We first neglect $h_{ij}$ to determine the profile of vector star and later take it into account to discuss the quantum decay. 
In particular, we have 
\beq
\sqrt{-g}=1+(3\Phi-\Psi)\,,
\eeq
for $h_{ij} = 0$.

\subsection{Configuration of vector boson star}

The Lorentz constraint \eq{eq:Lorentz1} is written as 
\beq
\label{eq:LorenzLinear}
\partial_0 A^0+\partial_i A^i +A^0\partial_0\left(3\Phi-\Psi\right)
+A^j\partial_j(3\Phi-\Psi)=0\,.
\eeq
We are interested in the mode that oscillates with a frequency of $\omega \simeq m$. 
This implies $A^0 = \mathcal{O}(A^i/(mR))$ at the leading order, where $R$ is the typical size of the object. 
For $mR \gg 1$, the Maxwell equations and the Lorentz condition are greatly simplified such as 
\beq
&&\partial_0^2 A^j-\Delta A^j
+\partial_0\left(5\Phi+\Psi\right)\partial_0 A^j
+\left[(1-2\Psi)m^2 + 2\partial_0^2 \Phi \right] A^j \simeq 0\,,\label{eom_Aj}
\\
&& \del_0 A^0+\partial_i A^i \simeq 0\,, 
\label{eq:Lorentz}
\eeq
where $\Delta=\partial_i\partial_j\delta^{ij}$. Here we included the leading-order terms that has time derivatives of gravitational potentials, which are negligible compared with the other terms, though. 
The Einstein equation leads to 
\beq
&&- \Delta \Phi=\frac{1}{2M_{\rm Pl}^2} T^0{}_0\,,\label{eq:DeltaPhi}\\
&&-\frac{2}{3}\Delta^2(\Psi-\Phi)=\frac{1}{M_{\rm Pl}^2}\left[\delta^{il}\delta^{jk}\partial_i\partial_j -\frac{\delta^{kl}\Delta}{3}\right] T^k{}_l\,,\label{eq:DeltaPsi}\\
&&\Delta V_i=\frac{2}{M_{\rm Pl}^2} T^{0}{}_i^{\rm (T)}\,,\\
&&-\partial_i \dot\Phi = \frac{1}{2M_{\rm Pl}^2} T^{0}{}_i^{\rm (L)}\,, \label{eq:didotPhi}
\eeq
where the superscript (T) and (L) indicate the transverse and longitudinal component, respectively.
Here, the $(0,0)$ and $(0,i)$ component of the energy-momentum tensor are given by 
\beq
 &&T^0{}_0 \simeq 
 \frac{m^2}{2} \lmk A^0 \rmk ^2 + 
 \frac{m^2}{2} \sum_i \lmk A^i \rmk^2 (1+2\Phi)
 + \frac{1}{2} \sum_i ( \del_0 A^i)^2 (1+2\Phi + 2\Psi)
 \nonumber\\
 && ~~~~~~~~~~~ 
 + \sum_i \del_0 A^i \del_i A^0 
 + \frac{1}{2} 
 \sum_{i,j} \lmk \del_i A^j \del_i A^j 
 - \del_i A^j \del_j A^i \rmk,  
 \label{eq:T00-2} \\
 &&T^0{}_i \simeq - m^2 A^0 A^i - 
 \sum_j \partial^0 A^j (\partial_j A^i - \partial_i A^j)\,,
 \label{eq:T0i}
\eeq
where we include terms up to $\mathcal{O}(k^2 (A^i)^2)$ 
for $T^0{}_0$ and leading-order terms for $T^0{}_i$.

Now, let us suppose 
\beq
 A^i = n^i f(r) \cos (\omega t + \theta^i)\,, 
 \label{eq:anzatzA}
\eeq
where $n_i$ is a unit vector representing the polarization and $\theta^i$ is the phase. Then, at the leading order, the equation of motion for $f$ is independent of the polarization and is given by 
\begin{align}
	\frac{\del^2 f}{\del r^2} + \frac{2}{r}\frac{\del f}{\del r}
         		- \lkk m^2 - \omega^2 (1+2 \Psi_0 (r) ) \rkk f =0\,, 
         		\label{eq:fr}
\end{align}
where $\Psi_0 (r)$ is the leading-order term of $\Psi$ and is given by 
\beq
\frac{1}{4\pi G} \left(\frac{\del^2  \Psi_0 }{\del r^2} + \frac{2}{r}\frac{\del  \Psi_0  }{\del r}\right)= - \frac{1}{2} m^2 f^2 (r)\,. 
\label{eq:fPsi}
\eeq
The latter equation comes from \eq{eq:DeltaPhi} by noting $\Psi_0 \simeq \Phi_0$ from \eq{eq:DeltaPsi} at the leading order.
These are identical to the equations for the scalar fields (see Eqs.~(\ref{eq:phir}) and (\ref{eq:Psi})). So the mass, size, and amplitude at the center of the vector boson star are the same with those for scalar boson star. 

%%%%%%%%%%%%%%%%%%%%%%%
\subsection{Oscillating behavior of the gravitational potential}
%%%%%%%%%%%%%%%%%%%%%%%

Similarly to the case of scalar boson star studied in the previous section, we are interested in the behavior of the gravitational potential $\Psi$ and $\Phi$ and its time dependence since it is essential to evaluate the quantum decay rate.
Compared with the scalar case, it is nontrivial since there are four components $A_\mu$ and several configurations of the vector polarization are known~\cite{Adshead:2021kvl,Jain:2021pnk,Zhang:2021xxa}.\footnote{
	Recall that the gravitational potential is inevitably oscillating for a real scalar case, but it can be time-independent for a complex scalar as shown in Sec.~\ref{sec:Q-ball}. 
}
Our short conclusion is that in the massive vector case, at least for known configurations, the gravitational potential oscillates with a similar amplitude. In deriving it, $A^0$ plays an important role since it is related to $A^i$ through the Lorentz constraint (\ref{eq:Lorentz}).

Using the ansatz \eq{eq:anzatzA}, 
we obtain the constraint from the Lorentz condition \eq{eq:Lorentz} such as 
\beq
 A^0 = - \frac{1}{\omega} \sum_i n^i \del_i f(r) \sin (\omega t + \theta^i) \,.
\eeq
Substituting these into $T^0{}_0$, 
we obtain 
\beq
 T^0{}_0 &&\simeq  
 \frac{m^2}{2\omega^2} \lmk \sum_i n^i \del_i f \sin (\omega t + \theta^i) \rmk ^2 + 
 \frac{m^2}{2} \sum_i \lmk n^i f(r) \cos (\omega t + \theta^i) \rmk^2 (1+2\Phi)
 \nonumber\\
 &&+ \frac{\omega^2}{2} \sum_i ( n^i f(r) \sin (\omega t + \theta^i))^2 (1+2\Phi + 2\Psi)
  \nonumber\\
 &&+ \sum_i n^i f(r) \sin (\omega t + \theta^i)
 \lmk \sum_j n^j \del_i \del_j f \sin(\omega t + \theta^j) \rmk 
 \nonumber\\
 &&+ \frac12 \sum_{i,j} \lmk n^j n^j \del_i f \del_i f \cos^2(\omega t + \theta^j) - n^i n^j \del_i f \del_j f \cos (\omega t + \theta^i) \cos(\omega t + \theta^j) \rmk
 \,.
  \nonumber\\
\eeq

In the following, we consider two extremely polarized configurations to see the time-dependence of gravitational potential.

%%%%%%%%%%%%%%%%%%%%%%%
\subsubsection{Linear polarization}
%%%%%%%%%%%%%%%%%%%%%%%

Let us consider the case with a linear polarization: 
$n^1 = 1$, $n^2 = 0$, $n^3 = 0$, 
$\theta^i = 0$. 
We then obtain 
\beq
 T^0{}_0 &&\simeq  
 \frac{1}{2} \hat{x}^2 (f')^2 \sin^2 (\omega t ) + 
 \frac{m^2}{2} (1+2\Phi) f^2 
 + \frac{(\omega^2 - m^2) + 2 \omega^2 \Psi}{2} f^2 \sin^2 (\omega t)
 \nonumber\\
 && + f 
 \lmk f'/r + \hat{x}^2 (f'' - f'/r) \rmk \sin^2 (\omega t) 
 + \frac12 f'^2 ( 1 - \hat{x}^2) \cos^2 (\omega t) 
 \,, 
 \nonumber\\
 &&= 
 \frac{m^2}{2} (1+2\Phi) f^2  
 + \lmk 
   - \frac{ff''}{2} 
 + \hat{x}^2 \lmk 
  f f''+ 
  \frac{f' f'}{2}
 - \frac{f f'}{r}
 \rmk \rmk \sin^2 (\omega t) 
 \nonumber\\
 && + \frac12 f'^2 ( 1 - \hat{x}^2) \cos^2 (\omega t)  
\,, 
\eeq
where $\hat{x} \equiv x / r$, the prime denotes the derivative with respect to $r$
and we used \eq{eq:fr} in the last line. 
The large parenthesis in front of $\sin^2(\omega t)$ is given by $\sim f^2/R^2$ with $R$ being the typical size of the vector boson star.
Noting that 
\beq
 \Delta f^2(r) = 2 \lmk ff'' + f'^2 +  \frac{2 ff'}{r} \rmk \sim \frac{f^2}{R^2} \,,
\eeq
and Eq.~(\ref{eq:DeltaPhi}), 
we approximate the gravitational potential as 
\begin{align}
	\Phi(t,r) \sim \Phi_0(r) + \frac{f^2}{16M_{\rm Pl}^2} \cos(2\omega t)\,, 
\end{align}
which should be regarded as an order-of-magnitude estimate. 
Here, we particularly omit the deviation from the spherical symmetry.

%%%%%%%%%%%%%%%%%%%%%%%
\subsubsection{Circular polarization}
%%%%%%%%%%%%%%%%%%%%%%%

Let us consider the case with a maximally circular polarization: 
$n^1 = 1/\sqrt{2}$, $n^2 = 1/\sqrt{2}$, $n^3 = 0$, 
$\theta^1 = 0$, $\theta^2 = \mp \pi/2$, $\theta^3 = 0$. 
We first note that $\sum_i n^i \del_i f \sin (\omega t + \theta^i) = f' (\hat{x} \sin (\omega t) \mp \hat{y} \cos(\omega t))/\sqrt{2} = f' \sin(\omega t \mp \phi_x) 
\sin \phi_z / \sqrt{2}$, where $x = r \cos \phi_x \sin \phi_z$ and $y = r \sin \phi_x \sin \phi_z$. 
We then obtain 
\beq
 T^0{}_0 &&\simeq  
 \frac{m^2}{2} (1 + 2\Phi) f^2 + 
 \frac{(\omega^2 - m^2) + 2 \omega^2 \Psi}{4} f^2 
 + \frac{1}{4} f' f' 
 \nonumber\\
 &&
 + \frac{1}{4} \sin^2 \phi_z f'f' \lmk \sin^2 (\omega t \mp \phi_x)  - \cos^2 (\omega t \mp \phi_x)  \rmk
 \nonumber\\
 &&
 + \frac{1}{2} f 
 \lkk \del_x^2 f \sin^2 (\omega t)  \mp 2 \del_x \del_y f \cos (\omega t) \sin (\omega t) + \del_y^2 f \cos^2 ( \omega t) \rkk. 
\eeq
Here, the third line can be rewritten as 
\beq
\frac{1}{2}\lkk 
\frac{ff'}{r} + \lmk f f'' - \frac{ff'}{r} \rmk \sin^2 (\omega t \mp \phi_x) \sin^2 \phi_z \rkk. 
\eeq
Using \eq{eq:fr}, 
we finally obtain 
\beq
 T^0{}_0 &&\simeq  
 \frac{m^2}{2} (1 + 2\Phi) f^2 
 + 
 \frac{1}{4} \lmk f' f' - f f'' \rmk 
 - \frac{1}{4} f' f' \cos^2 (\omega t \mp \phi_x) \sin^2 \phi_z
 \nonumber\\
 && 
 + \frac{1}{2} \lmk f f'' + \frac{1}{2} f' f' - \frac{ff'}{r} \rmk \sin^2 (\omega t \mp \phi_x) \sin^2 \phi_z 
  \,. 
\eeq
Thus we have the oscillating term.\footnote{
	Recently Ref.~\cite{Amin:2023imi} studied the real photon emission due to the higher dimensional interactions between the photon and dark photon. It has been found that some of the operators do not contribute to the photon emission for circular polarized vector solitons, since $|A_i|^2={\rm const.}$, neglecting the small $A_0$ contribution. For the graviton emission case, we do not have a freedom to choose such effective operators and the inclusion of the $A_0$ contribution is important.
}
Again, from Eq.~(\ref{eq:DeltaPhi}), we approximate the gravitational potential as
\begin{align}
	\Phi(t,r) \sim \Phi_0(r) + \frac{f^2}{16M_{\rm Pl}^2} \cos(2\omega t\pm 2\phi_x)\,.
\end{align}
This should be regarded as an order-of-magnitude estimate because we particularly omit the deviation from the spherical symmetry. 
This is the same orders-of-magnitude result as the linear polarized case.

%%%%%%%%%%%%%%%%%%%%%%%%%%%%%%%%
\section{Quantum decay into gravitons} \label{sec:decay}
%%%%%%%%%%%%%%%%%%%%%%%%%%%%%%%%

Now let us estimate the quantum production rate of the graviton in the boson star.
As far as we consider the graviton wavelength much shorter than the typical size of the boson star $R$, we can approximate the metric as if it is spatially flat.\footnote{
	See Ref.~\cite{Kawasaki:2013awa} for more precise discussion about this approximation for the case of scalar particle production.
} Then the the equation of motion of the gravitational wave $h_\lambda$, with $\lambda=+,\times$ being the polarization, is given by
\begin{align}
	\ddot h_{\lambda} + (3\dot\Phi+\dot\Psi) \dot h_\lambda + e^{-2(\Psi+\Phi)} k^2 h_\lambda = 0\,,  \label{GWEOM}
\end{align}
where $k$ denotes the wavenumber. 
One may think that both the oscillating $\Psi$ and $\Phi$ may induce the particle production, but we need to take care of the $e^{-2(\Psi+\Phi)}$ factor in front of the $k^2$ term. Actually it turns out that oscillation of $\Psi$ does not lead to particle production.
In order to see this, let us define a new time variable $d\tau \equiv e^{-(\Psi+\Phi)}dt$. Then the equation of motion (\ref{GWEOM}) is rewritten as
\begin{align}
	\partial^2_\tau h_{\lambda} + 2(\partial_\tau\Phi)(\partial_\tau h_\lambda) +k^2 h_\lambda = 0\,.
\end{align}
With this time variable, $\Psi$ disappeared and it is time dependence of $\Phi$ that leads to particle production.
This is further simplified by defining $\widetilde h_\lambda \equiv e^{\Phi} h_\lambda$ as\footnote{
	A new time variable is the analogue of the conformal time and $b$ is the analogue of the cosmic scale factor in the Freedman-Robertson-Walker universe.
}
\begin{align}
		\partial^2_\tau{\widetilde h}_{\lambda} + \left(k^2 -\frac{	\partial^2_\tau b}{b} \right) \widetilde h_\lambda = 0\,,  \label{h_eom}
\end{align}
where $b\equiv e^{\Phi}$. The time-dependent effective mass term is given by
\begin{align}
	&\frac{\partial^2_\tau b}{b} \simeq \left(\frac{\omega \phi_r}{2M_{\rm Pl}}\right)^2\cos(2\omega t) ~~~{\rm for~scalar}\,,\\
	&\frac{\partial^2_\tau b}{b} \simeq \left(\frac{\omega f}{2M_{\rm Pl}}\right)^2\cos(2\omega t) ~~~{\rm for~vector}\,,
\end{align}
from Eq.~(\ref{Phi_osc}) for the scalar case. 
Since the scalar and vector cases give similar results, below we collectively denote by $\phi$ both the scalar and vector boson for notational simplicity.

Equation~(\ref{h_eom}) is the same as the Mathieu equation and hence the modes with $k\simeq\omega$ are enhanced. 
Once the equation of motion is identified with the Mathieu equation, we can use the same technique to solve this equation (e.g. the Bogoliubov coefficient technique) and derive the production rate in the narrow width limit, as found in literature~\cite{Dolgov:1989us,Traschen:1990sw,Shtanov:1994ce,Kofman:1997yn}.
In the case of purely gravitational production, more details are found in Refs.~\cite{Ema:2015dka,Ema:2018ucl,Chung:2018ayg}.
Although we do not repeat the calculation details here, the point is that it is regarded as a perturbative annihilation of the scalar or vector into the graviton pair\footnote{
	One should be careful about this interpretation. It is justified after deriving the time dependence of $\Phi$, which exhibits $\cos(2\omega t)$ behavior. For the case of Q-ball, as shown in Sec.~\ref{sec:Q-ball}, the gravitational potential does not oscillate and this interpretation is not justified.   
} $\phi\phi\to hh$, since $(\phi_r/2M_{\rm Pl})^2, (f/2M_{\rm Pl})^2 \ll 1$.
The graviton production rate from a boson star is given by
\begin{align}
	\Gamma_{\rm grav}=\Gamma(\phi\phi\to hh) = \frac{\mathcal C}{256\pi} \frac{\phi_R^2 \ms^3}{M_{\rm Pl}^4}\,,
	\label{Gamma}
\end{align}
where $\phi_R$ denotes the typical amplitude of $\phi$ in the scalar boson star or $A_i$ in the vector boson star and we approximated $\omega \simeq \ms$, and $\mathcal C$ is an $\mathcal O(1)$ coefficient that represents the effect of radial shape of the boson star.

Note that this production mechanism is purely quantum. 
The production through (\ref{h_eom}) requires initial seed of the graviton field $\widetilde h_\lambda$, which is provided by the zero-point fluctuation in the vacuum. 
Thus it is not understood as a gravitational wave production from classical sources. 
Although a spherically symmetric boson star cannot be a classical source of gravitational waves, it can enhance the graviton quantum fluctuations existing there.

Here we comment on the possibility of parametric resonance. The equation of motion (\ref{h_eom}) is exactly the form of the Mathieu equation, and hence one may wonder whether the parametric resonant amplification of the graviton can happen or not. 
To see this, let us compare the growth rate of the graviton amplitude and the escape rate of the produced graviton from the soliton. If the former is larger than the latter, the parametric resonance can happen. Otherwise gravitons do not accumulate in the phase space with narrow momentum rage around $k\sim \omega$ and no Bose enhancement would be expected~\cite{Hertzberg:2010yz,Kawasaki:2013awa}.
The growth rate is characterized by the parameter $\mu$, which is defined as $h_\lambda \propto e^{\mu t}$. In our case it is given by $\mu \sim \omega\phi_R^2/M_{\rm Pl}^2$~\cite{Kofman:1997yn}. 
On the other hand, the escape rate is simply given by $\sim 1/R \sim \omega\phi_R/M_{\rm Pl}$ for the self-gravitating boson star, which is larger than the growth rate $\mu$. 
For the oscillon with self-interactions, the escape rate is even larger. 
Therefore, the escape rate is always larger than the growth rate for $\phi_R \ll M_{\rm Pl}$. 
Thus we conclude that there is no parametric resonance effect and the graviton production rate is linear in time with the perturbative rate (\ref{Gamma}).

All the Standard Model particles are also gravitationally coupled to $\phi$ and $A^\mu$ and hence their pair production can happen through the $s$-channel graviton exchange. Among them, massless fermion and transverse vector boson production are suppressed due to their conformal nature. The production rate of fermions is $(m_f / \ms)^2$ times smaller than the graviton production rate, where $m_f$ is the fermion mass. On the other hand, scalar production is not suppressed unless it is conformally coupled to the Ricci scalar. 
The equation of motion of a free scalar $\chi$, which is minimally coupled to gravity, is given by
\begin{align}
	e^{2\Psi}\left[\ddot\chi + (3\dot\Phi +\dot\Psi) \dot\chi\right] - e^{-2\Phi} \vec\nabla^2\chi + m_\chi^2\chi= 0 \,,
\end{align}
where $m_\chi$ is the mass of $\chi$. Similarly to the graviton case, it is simplified by introducing $\widetilde\chi\equiv e^{\Phi}\chi$ as
\begin{align}
		\partial^2_\tau\widetilde\chi +  \left(k^2 -\frac{\partial^2_\tau b}{b} + b^2 m_\chi^2\right)\widetilde\chi= 0\,,
\end{align}
where we have moved to the Fourier space. Therefore, if $k\simeq \ms\gg m_\chi$, the scalar production rate is the same as the graviton production rate up to a factor $2$ coming from the graviton polarization degrees of freedom.
In the Standard Model, the production rate of the Higgs boson is the same order of the graviton production rate if $\ms$ is much larger than the Higgs mass. In the electroweak symmetry breaking phase, the production of longitudinal massive vector bosons is also the same order.
If $0.1$\,GeV $\lesssim \ms \lesssim 1$\,GeV, the boson stars may also efficiently produce pions or other mesons.\footnote{
	The gravitational gluon production for $\ms \gtrsim 1\,{\rm GeV}$ may not be much suppressed compared with the graviton production due to the relatively strong breaking of scale invariance in the strong dynamics.
}
If there exist other scalar fields lighter than $\phi$ or $A^\mu$, such as axions, they should also be produced with the same production rate.

%%%%%%%%%%%%%%%%%%%%%%%%%%
\section{Phenomenological implications of gravitational decay}
\label{sec:application}
%%%%%%%%%%%%%%%%%%%%%%%%%%

In this section, we discuss time evolution of the boson star with a gravitational decay rate of (\ref{Gamma}) and then calculate the spectrum of graviton background. 
The following discussion applies to the case of both scalar and vector. For notational simplicity we simply express them by $\phi$, but it can be as either scalar or vector.

%%%%%%%%%%%%%%%%%%%%%%%%%%%%%%
\subsection{Decay of boson star} \label{sec:decay2}
%%%%%%%%%%%%%%%%%%%%%%%%%%%%%

If there are no direct interactions between $\phi$ (or $A^\mu$) and other lighter fields, the gravitational processes are the only processes that cause the decay of boson star.
Classical decay processes such as $3\phi\to\phi$ are possible mediated by the gravitational interaction~\cite{Page:2003rd}, but it is likely to be exponentially suppressed for $\omega R \gg 1$.
In this limit, the graviton production is an inevitable decay process and it gives a strict lower bound on the lifetime of oscillons/boson stars.

Via the graviton production, a boson star mass $M$ decreases according to
\begin{align}
	\dot M(t) = -\Gamma_{\rm grav}(t) M(t)\,.  \label{dotM}
\end{align}
Note that the decay rate (\ref{Gamma}) depends on the amplitude of the scalar/vector in the boson star $\phi_R$ and hence the decay rate decreases as the boson star mass decreases. 
By using the relation $M \sim \ms^2\phi_R^2 R^3$ and the equilibrium condition $GM \sim (R\ms^2)^{-1}$, we find $\Gamma_{\rm grav} \propto M^4$. Then we can solve (\ref{dotM}) as
\begin{align}
	M^4(t) = \frac{M_i^4}{1+ \Gamma_{\rm grav}^0 t},~~~~~~~~~\Gamma_{\rm grav}^0 \equiv  \frac{\mathcal C}{1024\pi} \frac{\phi_{R,i}^2 \ms^3}{M_{\rm Pl}^4}\,,
\end{align}
where we assumed the initial boson star mass $M_i$ in the limit $t\to 0$ and $\phi_{R,i}$ denotes the initial scalar/vector amplitude at the boson star formation. Correspondingly, we obtain
\begin{align}
	\phi_R^2(t) = \frac{\phi_{R,i}^2}{1+ \Gamma_{\rm grav}^0 t}\,.
\end{align}
Therefore, the boson star mass decreases rather slowly as $M \propto t^{-1/4}$ for $t \gtrsim (\Gamma_{\rm grav}^0)^{-1}$. We call $\tau_{\rm grav}\equiv (\Gamma_{\rm grav}^0)^{-1}$ as the boson star lifetime, although it does not decay exponentially. Numerically it is estimated as
\begin{align}
	\tau_{\rm grav}= (\Gamma_{\rm grav}^0)^{-1} \simeq
	1\times 10^{16}\,{\rm sec}\,\left(\frac{M_{\rm Pl}}{\phi_{R,i}}\right)^2 \left(\frac{1\,{\rm GeV}}{\ms}\right)^3\,.
	\label{tau_grav}
\end{align}
It easily exceeds the present age of the universe for, e.g., light axion-like particles $\ms \lesssim 1\,{\rm eV}$.
If $\ms > \mathcal O(100)$\,GeV, the boson stars may efficiently produce the Higgs boson and longitudinal weak bosons after the Big-Bang Nucleosynthesis epoch with a rate similar to the graviton production. This gives a constraint on the abundance of boson star.\footnote{
	If $0.1$\,GeV $\lesssim \ms \lesssim 1$\,GeV, the boson stars may also efficiently produce pions or other mesons after the recombination epoch, which would also give a tight constraint.
}
We will come back to this issue in detail elsewhere.

Once the amplitude $\phi_R(t)$ becomes smaller than $\ms$, the classical picture of the boson star may no longer be valid. This happens at $t \sim \left[\Gamma_{\rm grav}^0 (\ms/\phi_{R,i})^2 \right]^{-1}$. 
Then one has to treat the bosons as individual particles with negligible Bose enhancement effect.

%%%%%%%%%%%%%%%%%%%%%%%%%%%%%%
\subsection{Decay of oscillon and Q-ball}
%%%%%%%%%%%%%%%%%%%%%%%%%%%%%

Here we comment on the case with oscillon, where the self-interactions support the localized configuration. 
In most cases, oscillons classically decay into individual particles via self-interactions. 
In a certain case, however, 
the classical decay processes mediated by the self-interactions can be exponentially suppressed for $\omega R\gg 1$~\cite{Mukaida:2016hwd}.\footnote{
	For the Gaussian profile, the suppression factor should read $\sim \exp(-(\omega R)^2)$~\cite{Eby:2018ufi,Mukaida:2016hwd}. One should note that the decay rate is sensitive to the exact radial profile, since a small change in the radial profile may result in completely different functional form for its Fourier component, which is essential for the estimation of the classical decay rate. In this paper we do not go into details of this subject and just assume that it is exponentially suppressed for $\omega R\gg 1$.
}
In this limit, the graviton production is an inevitable decay process and it gives  strict lower bound on the lifetime of oscillon.

The graviton production rate is given by the same formula (\ref{Gamma}). In order to discuss the evolution of the oscillon, we need to know the relation between $\phi_R$ and $R$.
It is nontrivial and highly model-dependent. 
Nevertheless, we can make an orders-of-magnitude estimate for the typical time scale at which the gravitational process becomes important.
Independently of the relation between $\phi_R$ and $R$, we can repeat a similar procedure to the case of boson star, and find a typical time scale $\tau_{\rm grav}$, the same as (\ref{tau_grav}), although the time dependence of the boson star mass $M(t)$ at $t>\tau_{\rm grav}$ may differ from the boson star case.
Thus Eq.~(\ref{tau_grav}) gives a reasonable estimate for the oscillon lifetime for $\omega R\gg 1$ when the classical decay processes are exponentially suppressed.\footnote{
    For the so-called exact I-ball/oscillon~\cite{Kawasaki:2015vga}, there are no classical decay processes. Although fluctuations with particular modes may grow in analogy with the parametric resonance, still the final state is considered to be stable~\cite{Ibe:2019lzv}. In such a case, the graviton production is the only energy loss channel.
}

As we discussed in Sec.~\ref{sec:Q-ball}, 
the quantum decay of Q-ball into graviton is forbidden by the conserved global charge for the case with the exactly spherical orbit in the complex plane. 
If one adds higher-dimensional U(1)-breaking terms, 
the orbit in the complex plane becomes elliptical 
and the quantum decay process to the graviton becomes open. 
The amplitude of scalar field decreases as it decays, which makes decay slower. 
The resulting time-dependence for the graviton emission rate is different from the case with boson stars and provides an unique signals in gravitational wave spectrum.\footnote{
    The presence of U(1) breaking term itself tends to make Q-balls unstable. For the gravity mediation type Q-ball, for example, it has been shown that Q-balls decay within a simulation time if the magnitude of the U(1) breaking term is sizable~\cite{Kawasaki:2005xc}. However, still it is difficult to estimate the instability time scale for general (or much smaller) value of the U(1) breaking term. See also Refs.~\cite{Hiramatsu:2010dx,Hasegawa:2019bbo} for related works. 
    Moreover, dissipation via scatterings with thermal plasma makes the orbit in the complex plane being elliptical~\cite{Co:2019wyp,Co:2019jts}. This depends on the size of Q-balls and temperature of the ambient plasma. 
    Below we make an assumption that the elliptical orbit becomes circular only through the graviton emission. This gives a strict lower limit on the lifetime of the ``elliptical'' Q-ball, while it gives an upper bound on the graviton abundance. 
}

Below we estimate the effect of graviton production on the Q-ball. Let us suppose that the orbit of the complex scalar field $\phi$ is given by $\phi(t) = \varphi_R \cos(\omega t) + i\varphi_I \sin(\omega t)$ 
inside the Q-ball with $\varphi_R$ and $\varphi_I$ being the amplitude of the real and imaginary component.\footnote{
    The following estimate formally applies to the case of I-ball/oscillon by taking $\varphi_R \gg \varphi_I \to 0$.
} Without loss of generality, we take $\varphi_R > \varphi_I$ initially. From Eq.~(\ref{G01_complex}), the gravitational potential $\Phi(t,r)$ is given by
\begin{align}
	\Phi(t,r) \simeq \Phi_0(r) - \frac{\varphi_R^2(t,r)-\varphi_I^2(t,r)}{16 M_{\rm Pl}^2}\cos(2\omega t).
\end{align}
Repeating the same discussion for the case of real scalar, we obtain the graviton production rate due to the oscillating gravitational potential as
\begin{align}
	\Gamma_{\rm grav} = \frac{\mathcal C}{256\pi} \frac{ (\varphi_R^2-\varphi_I^2)\,\omega^3}{M_{\rm Pl}^4}.
\end{align}
It is expected that $\varphi_R^2-\varphi_I^2$ decreases through the graviton production and this process stops when $\varphi_R = \varphi_I$.
On the other hand, the Q-ball charge $Q$ should be conserved as far as there is no explicit U(1) breaking term. It is roughly estimated as $Q \sim \omega \varphi_R\varphi_I\times R^3 \sim \varphi_R\varphi_I/(\omega^2 K^3)$ where we assumed that the Q-ball radius is given by $R \sim 1/(\omega K)$ with some numerical constant $K$. 
Thus the final value will be $\varphi_R = \varphi_I \sim \omega \sqrt{K^3 Q} \equiv \bar\varphi$.
Keeping this in mind, one can solve (\ref{dotM}) in the case of Q-ball.

For $\varphi_R \gg \varphi_I$, the situation is similar to the case of real scalar. We obtain
\begin{align}
	\varphi_R^2(t) \simeq \frac{\varphi_{R,i}^2}{1+\Gamma^0_{\rm grav} t},~~~~~~M(t) \simeq \frac{M_{i}}{1+\Gamma^0_{\rm grav} t},~~~~~~
	\Gamma^0_{\rm grav} = \frac{\mathcal C}{256\pi} \frac{ \varphi_{R,i}^2\,\omega^3}{M_{\rm Pl}^4}.
\end{align}
Note that we have assumed that the Q-ball mass is given by $M \simeq \omega^2(\varphi_R^2+\varphi_I^2)\times R^3 \sim (\varphi_R^2+\varphi_I^2)/(\omega K^3)$. 
The amplitude $\varphi_R$ decreases according to this result. Eventually $\varphi_R$ becomes close to $\varphi_I$ at $t=\bar t$ and the above approximation breaks down. After this epoch, by defining $\varphi_R = \bar\varphi + \delta\varphi$ and $\varphi_I = \bar\varphi - \delta\varphi$ 
and noting that $M \sim (\bar\varphi^2+ \delta\varphi^2)/(\omega K^3)$, we obtain
\begin{align}
	\delta\varphi(t) = \delta\varphi(\bar t) - 2\overline\Gamma_{\rm grav} \bar\varphi(t-\bar t),~~~~~~M(t)\simeq M(\bar t),~~~~~~
	\overline\Gamma_{\rm grav} = \frac{\mathcal C}{256\pi} \frac{ \bar\varphi^2\,\omega^3}{M_{\rm Pl}^4} \sim \bar t^{-1}.
\end{align}
Note that $\overline\Gamma_{\rm grav}$ is defined by $\Gamma_{\rm grav} \simeq \overline\Gamma_{\rm grav} 4\delta\varphi/\bar\varphi$, so the graviton production vanishes for $\delta\varphi \to 0$.
Since $\delta\varphi(\bar t) \sim \bar\varphi$, we can easily see that $\delta\varphi (t)$ becomes zero after the time $t \sim \overline\Gamma_{\rm grav}^{-1}$ and the graviton production stops thereafter.
The above results are approximately combined in a convenient way as 
\begin{align}
	M(t) \simeq M_i \frac{1 + \overline\Gamma_{\rm grav} t}{1+ \Gamma_{\rm grav}^0 t},~~~~~~
	\Gamma_{\rm grav}(t) \simeq  \Gamma_{\rm grav}^0 \frac{1-  \overline\Gamma_{\rm grav} t}{1+ \Gamma_{\rm grav}^0 t},
\end{align}
for $t < \overline\Gamma_{\rm grav}^{-1}$.

%%%%%%%%%%%%%%%%%%%%%%%%%%%%%%%%
\subsection{Graviton background}
%%%%%%%%%%%%%%%%%%%%%%%%%%%%%%%%

One of the immediate consequences of the gravitational soliton decay is the formation of cosmic graviton background with a characteristic frequency spectrum.\footnote{Gravitational waves from the density perturbation during the formation of solitons such as oscillons~\cite{Zhou:2013tsa,Antusch:2016con,Liu:2017hua,Lozanov:2017hjm,Amin:2018xfe,Kitajima:2018zco,Liu:2018rrt,Lozanov:2019ylm,Hiramatsu:2020obh,Kou:2021bij,Garcia:2023eol} and Q-balls~\cite{Kusenko:2008zm,Kusenko:2009cv,Chiba:2009zu} have been studied. Our graviton production mechanism is completely different from them.} Even if they do not completely decay within the age of the universe due to the relatively slow decrease of the soliton mass $M$, the most fraction of their energy goes into the graviton radiation within the time scale of $\Gamma_{\rm grav}$. 
The present energy density of the graviton per logarithmic energy is calculated as
\begin{align}
	\frac{d\rho_{\rm GW}}{d\ln E} &= E^2\int dz \frac{N(z) n_{\rm sol}(z) a^3(z) \Gamma_{\rm grav}(z)}{H(z)} \frac{dN_{h}}{dE'}\\
	&= \frac{2E^4}{\ms^4} \frac{ \Gamma_{\rm grav}(z_E) \rho_{\rm sol}(z_E)}{H(z_E)}\,,
\end{align}
where $z$ denotes the redshift, $N(z)=M(z)/\ms$ is the number of $\phi$ in the soliton, $n_{\rm sol}(z)$ is the soliton number density, $a(z)=(1+z)^{-1}$ is the cosmic scale factor, $H(z)$ is the Hubble constant, $E'=(1+z)E$ and we approximate the produced graviton spectrum as a line: $dN_h/dE'=2\delta(E'-\ms)$.
In the second line we have performed the integral and $z_E=\ms/E-1$ and defined the soliton energy density $\rho_{\rm sol}(z)=M(z)n_{\rm sol}(z)$.

\begin{figure}[t]
  \centering
  \begin{tabular}{cc}
    \includegraphics[width=0.5\hsize]{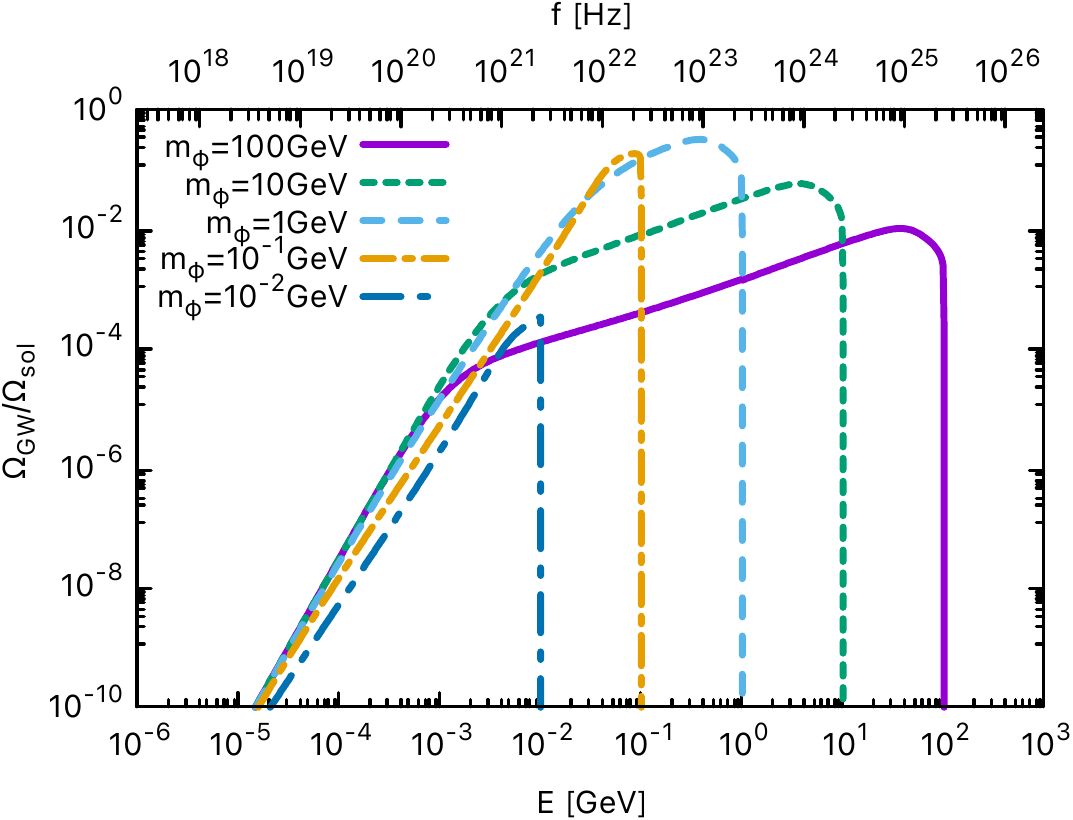}
     \includegraphics[width=0.5\hsize]{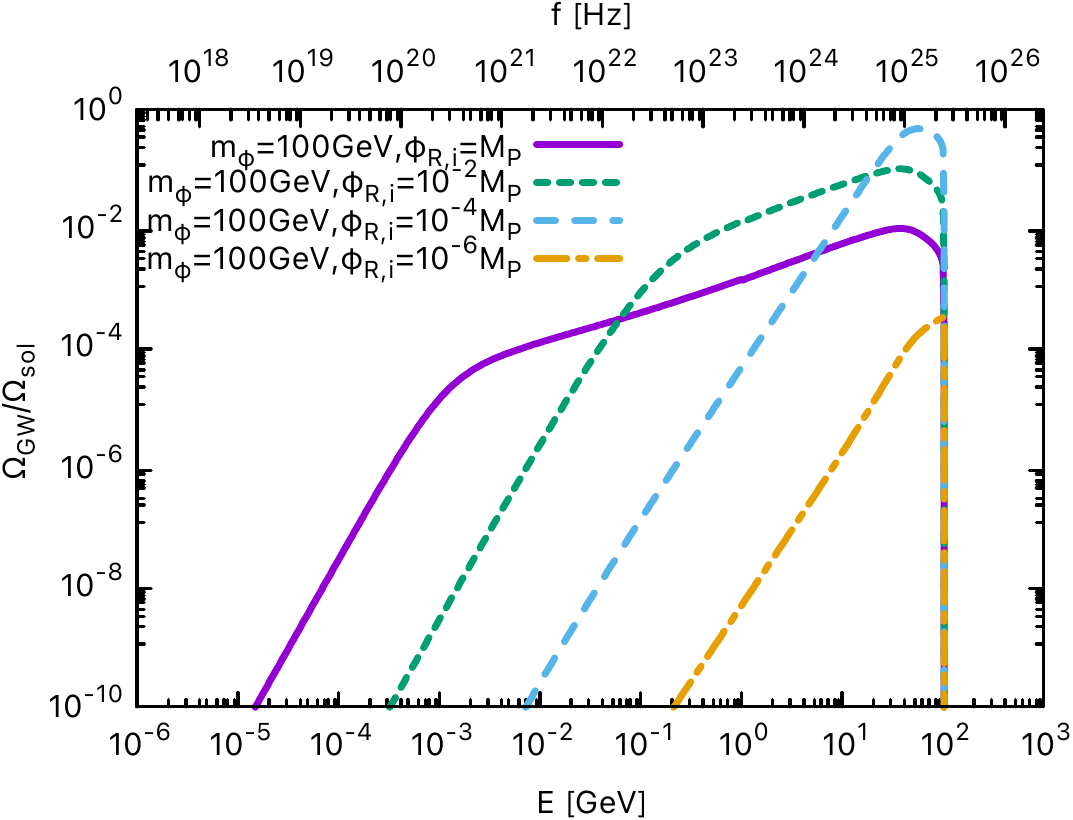}
  \end{tabular}
  \caption{(Left) The present graviton spectrum from the boson star decay for $\phi_{R,i}= M_{\rm Pl}$ with $\ms=100,10,1,10^{-1},10^{-2}\,{\rm GeV}$. (Right) The spectrum for $\ms=100\,{\rm GeV}$ with $\phi_{R,i}= (1,10^{-2},10^{-4},10^{-6})\times M_{\rm Pl}$. We are assuming that the boson star is long-lived and the dominant part comes from the boson star decaying at the present universe.}
  \label{fig:OGW}
\end{figure}

Fig.~\ref{fig:OGW} shows the graviton spectrum in the present universe from the decay of boson stars, assuming that the boson stars formed in the early universe lose their energy only through the graviton production.\footnote{
    It may be true for boson stars without self-interactions or for the exact I-balls/oscillons~\cite{Kawasaki:2015vga}.
}
The spectrum is represented in terms of $\Omega_{\rm GW} \equiv \frac{d\rho_{\rm GW}}{d\ln E} / \rho_{\rm crit}$ with $ \rho_{\rm crit}$ being the present critical density of the universe, normalized by the soliton energy density parameter $\Omega_{\rm sol}$. The soliton energy density parameter is defined by $\Omega_{\rm sol}\equiv \rho_{\rm sol}/\rho_{\rm crit}$ with neglecting the effect of decay.
In the left panel, the five lines show the case of $\ms=100,10,1,10^{-1},10^{-2}\,{\rm GeV}$ with $\phi_{R,i}= M_{\rm Pl}$.
Looking at the line of $m=100\,{\rm GeV}$, for example, the low frequency part corresponds to the gravitons produced at an early epoch when $ t \Gamma_{\rm grav}^0 < 1$, which exhibits the scaling $\Omega_{\rm GW}\propto E^3$. The middle part corresponds to those produced when $ t \Gamma_{\rm grav}^0 > 1$. Since the decay is not exponential but power-law as shown in Sec.~\ref{sec:decay2}, we obtain a spectrum with a power-law exponent different from the low-frequency part. It behaves as $\Omega_{\rm GW}\propto E^{5/8}$ in the matter-dominated universe and $\Omega_{\rm GW}\propto E^{1/2}$ in the radiation-dominated universe at the time of production.
Finally there is a cutoff at $E=\ms$ and a softening of the spectrum around the cutoff frequency is caused by the effect of cosmological constant.
These are unique features of the gravitational wave spectrum from the solitons, which is distinguished from, for example, a homogeneous scalar field decaying into two gravitons~\cite{Ema:2021fdz}.
The right panel shows the $\phi_{R,i}$ dependence on the spectrum for $m=100\,{\rm GeV}$. As is clear from Eq.~(\ref{tau_grav}), the peak amplitude is maximized around $\phi_{R,i}=10^{-4}M_{\rm Pl}$ since $\tau_{\rm grav}$ becomes close to the present age of the universe. For smaller $\phi_{R,i}$, $\tau_{\rm grav}$ becomes even longer and the graviton production efficiency is suppressed.

\begin{figure}[t]
  \centering
  \begin{tabular}{cc}
    \includegraphics[width=0.5\hsize]{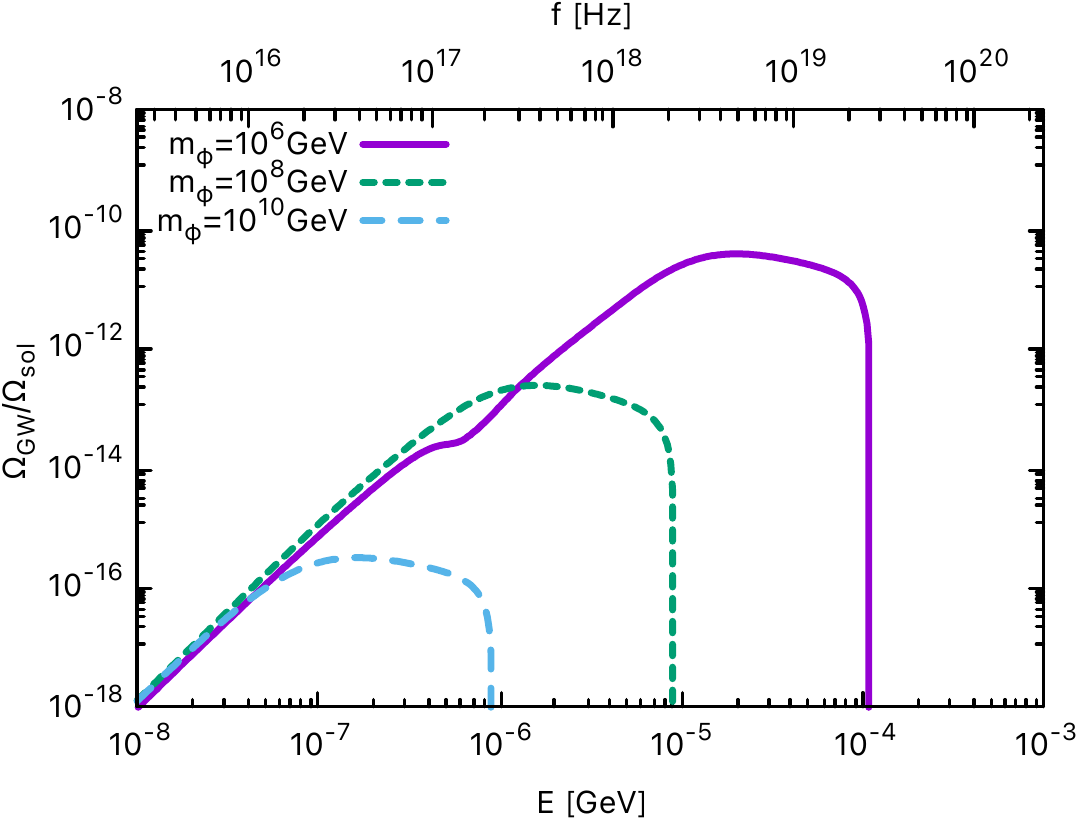}
     \includegraphics[width=0.5\hsize]{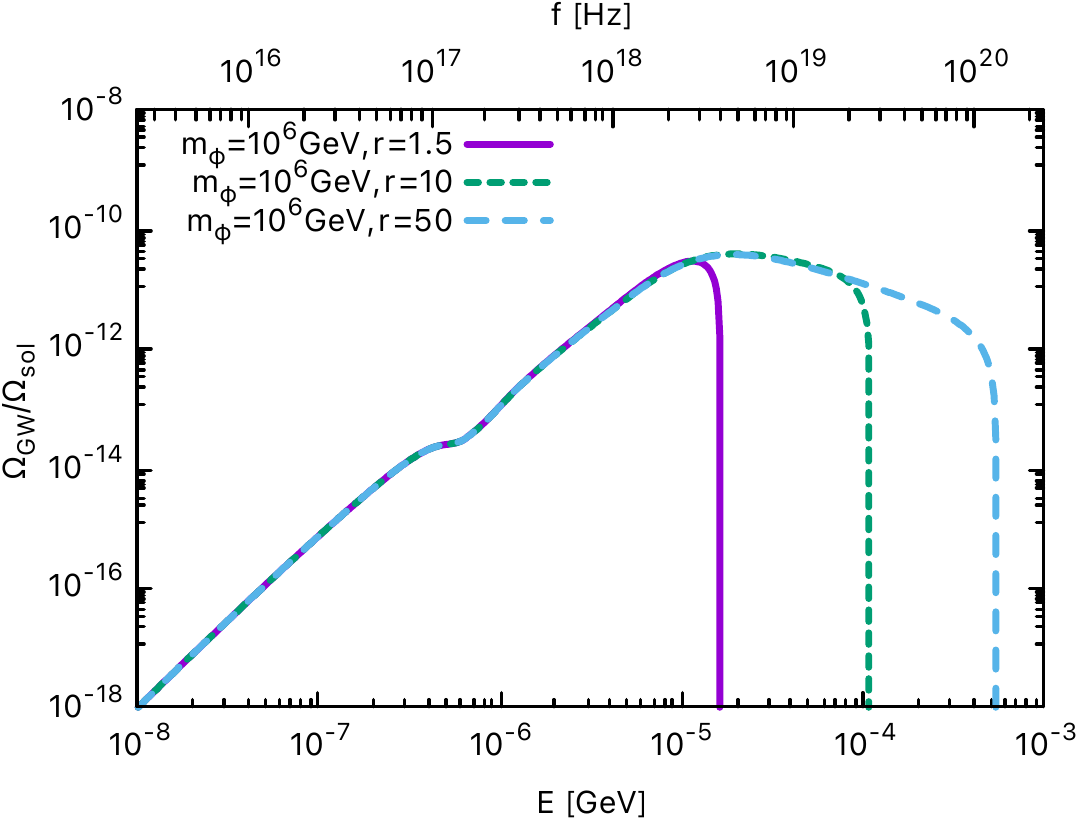}
  \end{tabular}
  \caption{The present stochastic gravitational wave spectrum from the Q-ball decay. (Left) We have taken $\ms=10^6, 10^8, 10^{10}\,{\rm GeV}$ with $\varphi_{R,i}= M_{\rm Pl}$ and $\bar\varphi = 0.1M_{\rm Pl}$, i.e., $r\equiv \varphi_{R,i} / \bar\varphi = 10$. (Right) We have taken $\ms=10^6\,{\rm GeV}$ and $r=1.5, 10, 50$ with $\varphi_{R,i}= M_{\rm Pl}$ fixed. The graviton production stops in the early universe when the orbit of the complex scalar becomes circular and the cutoff frequency roughly corresponds to the gravitons produced at this epoch.}
  \label{fig:OGWQ}
\end{figure}

Fig.~\ref{fig:OGWQ} shows the graviton spectrum in the present universe from the Q-ball decay. Note that similar results also apply to the case of oscillon, if the oscillon lifetime is long enough, as in the case of exact I-ball/oscillon. In the left panel we have taken $\ms=10^6, 10^8, 10^{10}\,{\rm GeV}$ with $\varphi_{R,i}= M_{\rm Pl}$ and $\bar\varphi = 0.1M_{\rm Pl}$, i.e., $r\equiv \varphi_{R,i} / \bar\varphi = 10$. In the right panel we have taken $\ms=10^6\,{\rm GeV}$ and $r=1.5, 10, 50$ with $\varphi_{R,i}= M_{\rm Pl}$ fixed. 
In this case, the graviton production stops in the early universe when the orbit of the complex scalar becomes circular and the cutoff frequency roughly corresponds to the gravitons produced at this epoch. 
Looking at the line of $r=50$ in the right panel, for example, the low frequency part corresponds to those produced at $t < (\Gamma_{\rm grav}^0)^{-1}$ that exhibits the scaling $\Omega_{\rm GW}\propto E^3$. A small modulation is caused by the change of relativistic degrees of freedom around the graviton emission epoch.
The higher frequency part corresponds to those produced at $(\Gamma_{\rm grav}^0)^{-1} < t < (\overline\Gamma_{\rm grav})^{-1}$ and the spectrum behaves as $\Omega_{\rm GW}\propto E^{-1}$. This is the characteristic signature of the gravitational wave spectrum from the Q-ball decay. 
The transition energy between these two regimes is given by $E\sim 10^{-6}\,{\rm GeV}\,(10^{10}\,{\rm GeV}/m)^{1/2} (M_{\rm Pl}/\varphi_{R,i})$.
Finally there is a cutoff around $E\sim 10^{-6}\,{\rm GeV}\,(10^{10}\,{\rm GeV}/m)^{1/2} (M_{\rm Pl}/\bar\varphi)$, corresponding to $t=(\overline\Gamma_{\rm grav})^{-1}$, after which the orbit of the complex scalar becomes completely circular and the graviton production stops.
For reference, this final epoch corresponds to the temperature $T \sim 10^4\,{\rm GeV}\,(m/10^{10}\,{\rm GeV})^{3/2} (\bar\varphi/M_{\rm Pl})$ if the universe is radiation-dominated.

In both cases, the predicted typical frequency of the stochastic gravitational wave is very high. Recently there are many proposals to detect such high frequency gravitational waves~\cite{Ejlli:2019bqj,Ringwald:2020ist,Aggarwal:2020olq,Berlin:2021txa,Domcke:2022rgu,Tobar:2022pie,Dolgov:2012be,Domcke:2020yzq,Ramazanov:2023nxz,Liu:2023mll,Ito:2023fcr}, although it is still challenging to detect it.

%%%%%%%%%%%%%%%%
%%%%%%%%%%%%%%%%

%%%%%%%%%%%%%%%%
%%%%%%%%%%%%%%%%

%%%%%%%%%%%%%%%%%%%%%%%%%%%%%%%%
\section{Discussion and conclusions}
\label{sec:conclusion}
%%%%%%%%%%%%%%%%%%%%%%%%%%%%%%%%

It is known that the cosmological dynamics of scalar or vector fields can lead to the formation of soliton-like objects such as an oscillon, Q-ball or boson stars. Oscillons/Q-balls are stabilized by self-interactions and boson stars are stabilized by gravitational interactions. They may have significant impacts on the early universe history, although still complete understandings of their dynamics have not been reached yet. One of the important properties of the soliton is its lifetime. There have been many studies for estimating the soliton lifetime, most of which focused on the emission of relativistic particles due to self-interactions.

In this paper we point out that there is a universal decay process of the solitons due to the gravitational interactions, i.e., quantum decay of a soliton into the graviton pair.
This decay channel exists even for a free (scalar or vector) field with a minimal coupling to gravity and provides a strict upper limit on their lifetime. 
Although the solitons are spherically symmetric, its gravitational potential has an oscillating component in time, which leads to the production of gravitational waves (gravitons) via quantum decay process. 
The quantum decay rate into gravitons has a power-law dependence on the size of solitons, which should be compared with the exponentially-suppressed classical decay rate. 
Therefore, it can be the dominant channel for boson stars consisting of fuzzy dark matter or vector dark matter without a kinetic mixing. 
We have discussed the evolution of solitons via the quantum decay 
and calculated the spectrum of stochastic gravitational waves. 
Although the amplitude of gravitational waves is too small to be detected in the near future, the spectrum has an unique shape  with a broken power law at a low frequency and a sharp cutoff at a high frequency.

%%%%%%%%%%%%%%%%%%%%%%%%%%%%%%%%
\section*{Acknowledgments}
This work was supported by MEXT Leading Initiative for Excellent Young Researchers (M.Y.), and by JSPS KAKENHI Grant No.\ 20H01894 (F.T.), 20H05851 (F.T. and M.Y.), and 23K13092 (M.Y.), and JSPS Core-to-Core Program (grant number: JPJSCCA20200002) (F.T.) and World Premier International Research Center Initiative (WPI), MEXT, Japan.
This article is based upon work from COST Action COSMIC WISPers CA21106, supported by COST (European Cooperation in Science and Technology).
%%%%%%%%%%%%%%%%%%%%%%%%%%%%%%%%

\bibliography{references}

\providecommand{\href}[2]{#2}\begingroup\raggedright\begin{thebibliography}{100}

\bibitem{Peccei:1977hh}
R.~D. Peccei and H.~R. Quinn, \emph{{CP Conservation in the Presence of
  Instantons}}, \href{https://doi.org/10.1103/PhysRevLett.38.1440}{\emph{Phys.
  Rev. Lett.} {\bfseries 38} (1977) 1440}.

\bibitem{Peccei:1977ur}
R.~D. Peccei and H.~R. Quinn, \emph{{Constraints Imposed by CP Conservation in
  the Presence of Instantons}},
  \href{https://doi.org/10.1103/PhysRevD.16.1791}{\emph{Phys. Rev. D}
  {\bfseries 16} (1977) 1791}.

\bibitem{Preskill:1982cy}
J.~Preskill, M.~B. Wise and F.~Wilczek, \emph{{Cosmology of the Invisible
  Axion}}, \href{https://doi.org/10.1016/0370-2693(83)90637-8}{\emph{Phys.
  Lett. B} {\bfseries 120} (1983) 127}.

\bibitem{Abbott:1982af}
L.~F. Abbott and P.~Sikivie, \emph{{A Cosmological Bound on the Invisible
  Axion}}, \href{https://doi.org/10.1016/0370-2693(83)90638-X}{\emph{Phys.
  Lett. B} {\bfseries 120} (1983) 133}.

\bibitem{Dine:1982ah}
M.~Dine and W.~Fischler, \emph{{The Not So Harmless Axion}},
  \href{https://doi.org/10.1016/0370-2693(83)90639-1}{\emph{Phys. Lett. B}
  {\bfseries 120} (1983) 137}.

\bibitem{Antypas:2022asj}
D.~Antypas et~al., \emph{{New Horizons: Scalar and Vector Ultralight Dark
  Matter}},  \href{https://arxiv.org/abs/2203.14915}{{\ttfamily 2203.14915}}.

\bibitem{Nelson:2011sf}
A.~E. Nelson and J.~Scholtz, \emph{{Dark Light, Dark Matter and the
  Misalignment Mechanism}},
  \href{https://doi.org/10.1103/PhysRevD.84.103501}{\emph{Phys. Rev. D}
  {\bfseries 84} (2011) 103501}
  [\href{https://arxiv.org/abs/1105.2812}{{\ttfamily 1105.2812}}].

\bibitem{Arias:2012az}
P.~Arias, D.~Cadamuro, M.~Goodsell, J.~Jaeckel, J.~Redondo and A.~Ringwald,
  \emph{{WISPy Cold Dark Matter}},
  \href{https://doi.org/10.1088/1475-7516/2012/06/013}{\emph{JCAP} {\bfseries
  06} (2012) 013} [\href{https://arxiv.org/abs/1201.5902}{{\ttfamily
  1201.5902}}].

\bibitem{Nakayama:2019rhg}
K.~Nakayama, \emph{{Vector Coherent Oscillation Dark Matter}},
  \href{https://doi.org/10.1088/1475-7516/2019/10/019}{\emph{JCAP} {\bfseries
  10} (2019) 019} [\href{https://arxiv.org/abs/1907.06243}{{\ttfamily
  1907.06243}}].

\bibitem{Nakayama:2020rka}
K.~Nakayama, \emph{{Constraint on Vector Coherent Oscillation Dark Matter with
  Kinetic Function}},
  \href{https://doi.org/10.1088/1475-7516/2020/08/033}{\emph{JCAP} {\bfseries
  08} (2020) 033} [\href{https://arxiv.org/abs/2004.10036}{{\ttfamily
  2004.10036}}].

\bibitem{Kitajima:2023fun}
N.~Kitajima and K.~Nakayama, \emph{{Viable Vector Coherent Oscillation Dark
  Matter}},  \href{https://arxiv.org/abs/2303.04287}{{\ttfamily 2303.04287}}.

\bibitem{Long:2019lwl}
A.~J. Long and L.-T. Wang, \emph{{Dark Photon Dark Matter from a Network of
  Cosmic Strings}},
  \href{https://doi.org/10.1103/PhysRevD.99.063529}{\emph{Phys. Rev. D}
  {\bfseries 99} (2019) 063529}
  [\href{https://arxiv.org/abs/1901.03312}{{\ttfamily 1901.03312}}].

\bibitem{Kitajima:2022lre}
N.~Kitajima and K.~Nakayama, \emph{{Dark Photon Dark Matter from Cosmic Strings
  and Gravitational Wave Background}},
  \href{https://arxiv.org/abs/2212.13573}{{\ttfamily 2212.13573}}.

\bibitem{Nakayama:2021avl}
K.~Nakayama and W.~Yin, \emph{{Hidden photon and axion dark matter from
  symmetry breaking}},
  \href{https://doi.org/10.1007/JHEP10(2021)026}{\emph{JHEP} {\bfseries 10}
  (2021) 026} [\href{https://arxiv.org/abs/2105.14549}{{\ttfamily
  2105.14549}}].

\bibitem{Salehian:2020asa}
B.~Salehian, M.~A. Gorji, H.~Firouzjahi and S.~Mukohyama, \emph{{Vector dark
  matter production from inflation with symmetry breaking}},
  \href{https://doi.org/10.1103/PhysRevD.103.063526}{\emph{Phys. Rev. D}
  {\bfseries 103} (2021) 063526}
  [\href{https://arxiv.org/abs/2010.04491}{{\ttfamily 2010.04491}}].

\bibitem{Firouzjahi:2020whk}
H.~Firouzjahi, M.~A. Gorji, S.~Mukohyama and B.~Salehian, \emph{{Dark photon
  dark matter from charged inflaton}},
  \href{https://doi.org/10.1007/JHEP06(2021)050}{\emph{JHEP} {\bfseries 06}
  (2021) 050} [\href{https://arxiv.org/abs/2011.06324}{{\ttfamily
  2011.06324}}].

\bibitem{Agrawal:2018vin}
P.~Agrawal, N.~Kitajima, M.~Reece, T.~Sekiguchi and F.~Takahashi, \emph{{Relic
  Abundance of Dark Photon Dark Matter}},
  \href{https://doi.org/10.1016/j.physletb.2019.135136}{\emph{Phys. Lett. B}
  {\bfseries 801} (2020) 135136}
  [\href{https://arxiv.org/abs/1810.07188}{{\ttfamily 1810.07188}}].

\bibitem{Dror:2018pdh}
J.~A. Dror, K.~Harigaya and V.~Narayan, \emph{{Parametric Resonance Production
  of Ultralight Vector Dark Matter}},
  \href{https://doi.org/10.1103/PhysRevD.99.035036}{\emph{Phys. Rev. D}
  {\bfseries 99} (2019) 035036}
  [\href{https://arxiv.org/abs/1810.07195}{{\ttfamily 1810.07195}}].

\bibitem{Co:2018lka}
R.~T. Co, A.~Pierce, Z.~Zhang and Y.~Zhao, \emph{{Dark Photon Dark Matter
  Produced by Axion Oscillations}},
  \href{https://doi.org/10.1103/PhysRevD.99.075002}{\emph{Phys. Rev. D}
  {\bfseries 99} (2019) 075002}
  [\href{https://arxiv.org/abs/1810.07196}{{\ttfamily 1810.07196}}].

\bibitem{Bastero-Gil:2018uel}
M.~Bastero-Gil, J.~Santiago, L.~Ubaldi and R.~Vega-Morales, \emph{{Vector dark
  matter production at the end of inflation}},
  \href{https://doi.org/10.1088/1475-7516/2019/04/015}{\emph{JCAP} {\bfseries
  04} (2019) 015} [\href{https://arxiv.org/abs/1810.07208}{{\ttfamily
  1810.07208}}].

\bibitem{Co:2021rhi}
R.~T. Co, K.~Harigaya and A.~Pierce, \emph{{Gravitational waves and dark photon
  dark matter from axion rotations}},
  \href{https://doi.org/10.1007/JHEP12(2021)099}{\emph{JHEP} {\bfseries 12}
  (2021) 099} [\href{https://arxiv.org/abs/2104.02077}{{\ttfamily
  2104.02077}}].

\bibitem{Kitajima:2023pby}
N.~Kitajima and F.~Takahashi, \emph{{Resonant production of dark photons from
  axion without a large coupling}},
  \href{https://arxiv.org/abs/2303.05492}{{\ttfamily 2303.05492}}.

\bibitem{Graham:2015rva}
P.~W. Graham, J.~Mardon and S.~Rajendran, \emph{{Vector Dark Matter from
  Inflationary Fluctuations}},
  \href{https://doi.org/10.1103/PhysRevD.93.103520}{\emph{Phys. Rev. D}
  {\bfseries 93} (2016) 103520}
  [\href{https://arxiv.org/abs/1504.02102}{{\ttfamily 1504.02102}}].

\bibitem{Ema:2019yrd}
Y.~Ema, K.~Nakayama and Y.~Tang, \emph{{Production of purely gravitational dark
  matter: the case of fermion and vector boson}},
  \href{https://doi.org/10.1007/JHEP07(2019)060}{\emph{JHEP} {\bfseries 07}
  (2019) 060} [\href{https://arxiv.org/abs/1903.10973}{{\ttfamily
  1903.10973}}].

\bibitem{Ahmed:2020fhc}
A.~Ahmed, B.~Grzadkowski and A.~Socha, \emph{{Gravitational production of
  vector dark matter}},
  \href{https://doi.org/10.1007/JHEP08(2020)059}{\emph{JHEP} {\bfseries 08}
  (2020) 059} [\href{https://arxiv.org/abs/2005.01766}{{\ttfamily
  2005.01766}}].

\bibitem{Nakayama:2020ikz}
K.~Nakayama and Y.~Tang, \emph{{Gravitational Production of Hidden Photon Dark
  Matter in Light of the XENON1T Excess}},
  \href{https://doi.org/10.1016/j.physletb.2020.135977}{\emph{Phys. Lett. B}
  {\bfseries 811} (2020) 135977}
  [\href{https://arxiv.org/abs/2006.13159}{{\ttfamily 2006.13159}}].

\bibitem{Kolb:2020fwh}
E.~W. Kolb and A.~J. Long, \emph{{Completely dark photons from gravitational
  particle production during the inflationary era}},
  \href{https://doi.org/10.1007/JHEP03(2021)283}{\emph{JHEP} {\bfseries 03}
  (2021) 283} [\href{https://arxiv.org/abs/2009.03828}{{\ttfamily
  2009.03828}}].

\bibitem{Arvanitaki:2021qlj}
A.~Arvanitaki, S.~Dimopoulos, M.~Galanis, D.~Racco, O.~Simon and J.~O.
  Thompson, \emph{{Dark QED from inflation}},
  \href{https://doi.org/10.1007/JHEP11(2021)106}{\emph{JHEP} {\bfseries 11}
  (2021) 106} [\href{https://arxiv.org/abs/2108.04823}{{\ttfamily
  2108.04823}}].

\bibitem{Sato:2022jya}
T.~Sato, F.~Takahashi and M.~Yamada, \emph{{Gravitational production of dark
  photon dark matter with mass generated by the Higgs mechanism}},
  \href{https://doi.org/10.1088/1475-7516/2022/08/022}{\emph{JCAP} {\bfseries
  08} (2022) 022} [\href{https://arxiv.org/abs/2204.11896}{{\ttfamily
  2204.11896}}].

\bibitem{Guth:2014hsa}
A.~H. Guth, M.~P. Hertzberg and C.~Prescod-Weinstein, \emph{{Do Dark Matter
  Axions Form a Condensate with Long-Range Correlation?}},
  \href{https://doi.org/10.1103/PhysRevD.92.103513}{\emph{Phys. Rev. D}
  {\bfseries 92} (2015) 103513}
  [\href{https://arxiv.org/abs/1412.5930}{{\ttfamily 1412.5930}}].

\bibitem{Kaup:1968zz}
D.~J. Kaup, \emph{{Klein-Gordon Geon}},
  \href{https://doi.org/10.1103/PhysRev.172.1331}{\emph{Phys. Rev.} {\bfseries
  172} (1968) 1331}.

\bibitem{Ruffini:1969qy}
R.~Ruffini and S.~Bonazzola, \emph{{Systems of selfgravitating particles in
  general relativity and the concept of an equation of state}},
  \href{https://doi.org/10.1103/PhysRev.187.1767}{\emph{Phys. Rev.} {\bfseries
  187} (1969) 1767}.

\bibitem{Colpi:1986ye}
M.~Colpi, S.~L. Shapiro and I.~Wasserman, \emph{{Boson Stars: Gravitational
  Equilibria of Selfinteracting Scalar Fields}},
  \href{https://doi.org/10.1103/PhysRevLett.57.2485}{\emph{Phys. Rev. Lett.}
  {\bfseries 57} (1986) 2485}.

\bibitem{Seidel:1991zh}
E.~Seidel and W.~M. Suen, \emph{{Oscillating soliton stars}},
  \href{https://doi.org/10.1103/PhysRevLett.66.1659}{\emph{Phys. Rev. Lett.}
  {\bfseries 66} (1991) 1659}.

\bibitem{Tkachev:1991ka}
I.~I. Tkachev, \emph{{On the possibility of Bose star formation}},
  \href{https://doi.org/10.1016/0370-2693(91)90330-S}{\emph{Phys. Lett. B}
  {\bfseries 261} (1991) 289}.

\bibitem{Kolb:1993zz}
E.~W. Kolb and I.~I. Tkachev, \emph{{Axion miniclusters and Bose stars}},
  \href{https://doi.org/10.1103/PhysRevLett.71.3051}{\emph{Phys. Rev. Lett.}
  {\bfseries 71} (1993) 3051}
  [\href{https://arxiv.org/abs/hep-ph/9303313}{{\ttfamily hep-ph/9303313}}].

\bibitem{Kolb:1993hw}
E.~W. Kolb and I.~I. Tkachev, \emph{{Nonlinear axion dynamics and formation of
  cosmological pseudosolitons}},
  \href{https://doi.org/10.1103/PhysRevD.49.5040}{\emph{Phys. Rev. D}
  {\bfseries 49} (1994) 5040}
  [\href{https://arxiv.org/abs/astro-ph/9311037}{{\ttfamily
  astro-ph/9311037}}].

\bibitem{Schive:2014hza}
H.-Y. Schive, M.-H. Liao, T.-P. Woo, S.-K. Wong, T.~Chiueh, T.~Broadhurst
  et~al., \emph{{Understanding the Core-Halo Relation of Quantum Wave Dark
  Matter from 3D Simulations}},
  \href{https://doi.org/10.1103/PhysRevLett.113.261302}{\emph{Phys. Rev. Lett.}
  {\bfseries 113} (2014) 261302}
  [\href{https://arxiv.org/abs/1407.7762}{{\ttfamily 1407.7762}}].

\bibitem{Levkov:2018kau}
D.~G. Levkov, A.~G. Panin and I.~I. Tkachev, \emph{{Gravitational Bose-Einstein
  condensation in the kinetic regime}},
  \href{https://doi.org/10.1103/PhysRevLett.121.151301}{\emph{Phys. Rev. Lett.}
  {\bfseries 121} (2018) 151301}
  [\href{https://arxiv.org/abs/1804.05857}{{\ttfamily 1804.05857}}].

\bibitem{Widdicombe:2018oeo}
J.~Y. Widdicombe, T.~Helfer, D.~J.~E. Marsh and E.~A. Lim, \emph{{Formation of
  Relativistic Axion Stars}},
  \href{https://doi.org/10.1088/1475-7516/2018/10/005}{\emph{JCAP} {\bfseries
  10} (2018) 005} [\href{https://arxiv.org/abs/1806.09367}{{\ttfamily
  1806.09367}}].

\bibitem{Eggemeier:2019jsu}
B.~Eggemeier and J.~C. Niemeyer, \emph{{Formation and mass growth of axion
  stars in axion miniclusters}},
  \href{https://doi.org/10.1103/PhysRevD.100.063528}{\emph{Phys. Rev. D}
  {\bfseries 100} (2019) 063528}
  [\href{https://arxiv.org/abs/1906.01348}{{\ttfamily 1906.01348}}].

\bibitem{Chen:2020cef}
J.~Chen, X.~Du, E.~W. Lentz, D.~J.~E. Marsh and J.~C. Niemeyer, \emph{{New
  insights into the formation and growth of boson stars in dark matter halos}},
  \href{https://doi.org/10.1103/PhysRevD.104.083022}{\emph{Phys. Rev. D}
  {\bfseries 104} (2021) 083022}
  [\href{https://arxiv.org/abs/2011.01333}{{\ttfamily 2011.01333}}].

\bibitem{Braaten:2015eeu}
E.~Braaten, A.~Mohapatra and H.~Zhang, \emph{{Dense Axion Stars}},
  \href{https://doi.org/10.1103/PhysRevLett.117.121801}{\emph{Phys. Rev. Lett.}
  {\bfseries 117} (2016) 121801}
  [\href{https://arxiv.org/abs/1512.00108}{{\ttfamily 1512.00108}}].

\bibitem{Braaten:2016dlp}
E.~Braaten, A.~Mohapatra and H.~Zhang, \emph{{Emission of Photons and
  Relativistic Axions from Axion Stars}},
  \href{https://doi.org/10.1103/PhysRevD.96.031901}{\emph{Phys. Rev. D}
  {\bfseries 96} (2017) 031901}
  [\href{https://arxiv.org/abs/1609.05182}{{\ttfamily 1609.05182}}].

\bibitem{Visinelli:2017ooc}
L.~Visinelli, S.~Baum, J.~Redondo, K.~Freese and F.~Wilczek, \emph{{Dilute and
  dense axion stars}},
  \href{https://doi.org/10.1016/j.physletb.2017.12.010}{\emph{Phys. Lett. B}
  {\bfseries 777} (2018) 64}
  [\href{https://arxiv.org/abs/1710.08910}{{\ttfamily 1710.08910}}].

\bibitem{Schiappacasse:2017ham}
E.~D. Schiappacasse and M.~P. Hertzberg, \emph{{Analysis of Dark Matter Axion
  Clumps with Spherical Symmetry}},
  \href{https://doi.org/10.1088/1475-7516/2018/01/037}{\emph{JCAP} {\bfseries
  01} (2018) 037} [\href{https://arxiv.org/abs/1710.04729}{{\ttfamily
  1710.04729}}].

\bibitem{Veltmaat:2018dfz}
J.~Veltmaat, J.~C. Niemeyer and B.~Schwabe, \emph{{Formation and structure of
  ultralight bosonic dark matter halos}},
  \href{https://doi.org/10.1103/PhysRevD.98.043509}{\emph{Phys. Rev. D}
  {\bfseries 98} (2018) 043509}
  [\href{https://arxiv.org/abs/1804.09647}{{\ttfamily 1804.09647}}].

\bibitem{Chavanis:2011zi}
P.-H. Chavanis, \emph{{Mass-radius relation of Newtonian self-gravitating
  Bose-Einstein condensates with short-range interactions: I. Analytical
  results}}, \href{https://doi.org/10.1103/PhysRevD.84.043531}{\emph{Phys. Rev.
  D} {\bfseries 84} (2011) 043531}
  [\href{https://arxiv.org/abs/1103.2050}{{\ttfamily 1103.2050}}].

\bibitem{Chavanis:2011zm}
P.~H. Chavanis and L.~Delfini, \emph{{Mass-radius relation of Newtonian
  self-gravitating Bose-Einstein condensates with short-range interactions: II.
  Numerical results}},
  \href{https://doi.org/10.1103/PhysRevD.84.043532}{\emph{Phys. Rev. D}
  {\bfseries 84} (2011) 043532}
  [\href{https://arxiv.org/abs/1103.2054}{{\ttfamily 1103.2054}}].

\bibitem{Eby:2018ufi}
J.~Eby, K.~Mukaida, M.~Takimoto, L.~C.~R. Wijewardhana and M.~Yamada,
  \emph{{Classical nonrelativistic effective field theory and the role of
  gravitational interactions}},
  \href{https://doi.org/10.1103/PhysRevD.99.123503}{\emph{Phys. Rev. D}
  {\bfseries 99} (2019) 123503}
  [\href{https://arxiv.org/abs/1807.09795}{{\ttfamily 1807.09795}}].

\bibitem{Mukaida:2016hwd}
K.~Mukaida, M.~Takimoto and M.~Yamada, \emph{{On Longevity of
  I-ball/Oscillon}}, \href{https://doi.org/10.1007/JHEP03(2017)122}{\emph{JHEP}
  {\bfseries 03} (2017) 122}
  [\href{https://arxiv.org/abs/1612.07750}{{\ttfamily 1612.07750}}].

\bibitem{Chan:2022bkz}
J.~H.-H. Chan, S.~Sibiryakov and W.~Xue, \emph{{Condensation and Evaporation of
  Boson Stars}},  \href{https://arxiv.org/abs/2207.04057}{{\ttfamily
  2207.04057}}.

\bibitem{Fujikura:2021omw}
K.~Fujikura, M.~P. Hertzberg, E.~D. Schiappacasse and M.~Yamaguchi,
  \emph{{Microlensing constraints on axion stars including finite lens and
  source size effects}},
  \href{https://doi.org/10.1103/PhysRevD.104.123012}{\emph{Phys. Rev. D}
  {\bfseries 104} (2021) 123012}
  [\href{https://arxiv.org/abs/2109.04283}{{\ttfamily 2109.04283}}].

\bibitem{Ellis:2022grh}
D.~Ellis, D.~J.~E. Marsh, B.~Eggemeier, J.~Niemeyer, J.~Redondo and K.~Dolag,
  \emph{{Structure of axion miniclusters}},
  \href{https://doi.org/10.1103/PhysRevD.106.103514}{\emph{Phys. Rev. D}
  {\bfseries 106} (2022) 103514}
  [\href{https://arxiv.org/abs/2204.13187}{{\ttfamily 2204.13187}}].

\bibitem{Bogolyubsky:1976nx}
I.~L. Bogolyubsky and V.~G. Makhankov, \emph{{On the Pulsed Soliton Lifetime in
  Two Classical Relativistic Theory Models}}, {\emph{JETP Lett.} {\bfseries 24}
  (1976) 12}.

\bibitem{Segur:1987mg}
H.~Segur and M.~D. Kruskal, \emph{{Nonexistence of Small Amplitude Breather
  Solutions in $\phi^4$ Theory}},
  \href{https://doi.org/10.1103/PhysRevLett.58.747}{\emph{Phys. Rev. Lett.}
  {\bfseries 58} (1987) 747}.

\bibitem{Gleiser:1993pt}
M.~Gleiser, \emph{{Pseudostable bubbles}},
  \href{https://doi.org/10.1103/PhysRevD.49.2978}{\emph{Phys. Rev. D}
  {\bfseries 49} (1994) 2978}
  [\href{https://arxiv.org/abs/hep-ph/9308279}{{\ttfamily hep-ph/9308279}}].

\bibitem{Copeland:1995fq}
E.~J. Copeland, M.~Gleiser and H.~R. Muller, \emph{{Oscillons: Resonant
  configurations during bubble collapse}},
  \href{https://doi.org/10.1103/PhysRevD.52.1920}{\emph{Phys. Rev. D}
  {\bfseries 52} (1995) 1920}
  [\href{https://arxiv.org/abs/hep-ph/9503217}{{\ttfamily hep-ph/9503217}}].

\bibitem{Gleiser:1999tj}
M.~Gleiser and A.~Sornborger, \emph{{Longlived localized field configurations
  in small lattices: Application to oscillons}},
  \href{https://doi.org/10.1103/PhysRevE.62.1368}{\emph{Phys. Rev. E}
  {\bfseries 62} (2000) 1368}
  [\href{https://arxiv.org/abs/patt-sol/9909002}{{\ttfamily
  patt-sol/9909002}}].

\bibitem{Honda:2001xg}
E.~P. Honda and M.~W. Choptuik, \emph{{Fine structure of oscillons in the
  spherically symmetric phi**4 Klein-Gordon model}},
  \href{https://doi.org/10.1103/PhysRevD.65.084037}{\emph{Phys. Rev. D}
  {\bfseries 65} (2002) 084037}
  [\href{https://arxiv.org/abs/hep-ph/0110065}{{\ttfamily hep-ph/0110065}}].

\bibitem{Kasuya:2002zs}
S.~Kasuya, M.~Kawasaki and F.~Takahashi, \emph{{I-balls}},
  \href{https://doi.org/10.1016/S0370-2693(03)00344-7}{\emph{Phys. Lett. B}
  {\bfseries 559} (2003) 99}
  [\href{https://arxiv.org/abs/hep-ph/0209358}{{\ttfamily hep-ph/0209358}}].

\bibitem{Fodor:2006zs}
G.~Fodor, P.~Forgacs, P.~Grandclement and I.~Racz, \emph{{Oscillons and
  Quasi-breathers in the phi**4 Klein-Gordon model}},
  \href{https://doi.org/10.1103/PhysRevD.74.124003}{\emph{Phys. Rev. D}
  {\bfseries 74} (2006) 124003}
  [\href{https://arxiv.org/abs/hep-th/0609023}{{\ttfamily hep-th/0609023}}].

\bibitem{Fodor:2008du}
G.~Fodor, P.~Forgacs, Z.~Horvath and M.~Mezei, \emph{{Computation of the
  radiation amplitude of oscillons}},
  \href{https://doi.org/10.1103/PhysRevD.79.065002}{\emph{Phys. Rev. D}
  {\bfseries 79} (2009) 065002}
  [\href{https://arxiv.org/abs/0812.1919}{{\ttfamily 0812.1919}}].

\bibitem{Gleiser:2008ty}
M.~Gleiser and D.~Sicilia, \emph{{Analytical Characterization of Oscillon
  Energy and Lifetime}},
  \href{https://doi.org/10.1103/PhysRevLett.101.011602}{\emph{Phys. Rev. Lett.}
  {\bfseries 101} (2008) 011602}
  [\href{https://arxiv.org/abs/0804.0791}{{\ttfamily 0804.0791}}].

\bibitem{Gleiser:2009ys}
M.~Gleiser and D.~Sicilia, \emph{{A General Theory of Oscillon Dynamics}},
  \href{https://doi.org/10.1103/PhysRevD.80.125037}{\emph{Phys. Rev. D}
  {\bfseries 80} (2009) 125037}
  [\href{https://arxiv.org/abs/0910.5922}{{\ttfamily 0910.5922}}].

\bibitem{Amin:2010jq}
M.~A. Amin and D.~Shirokoff, \emph{{Flat-top oscillons in an expanding
  universe}}, \href{https://doi.org/10.1103/PhysRevD.81.085045}{\emph{Phys.
  Rev. D} {\bfseries 81} (2010) 085045}
  [\href{https://arxiv.org/abs/1002.3380}{{\ttfamily 1002.3380}}].

\bibitem{Amin:2011hj}
M.~A. Amin, R.~Easther, H.~Finkel, R.~Flauger and M.~P. Hertzberg,
  \emph{{Oscillons After Inflation}},
  \href{https://doi.org/10.1103/PhysRevLett.108.241302}{\emph{Phys. Rev. Lett.}
  {\bfseries 108} (2012) 241302}
  [\href{https://arxiv.org/abs/1106.3335}{{\ttfamily 1106.3335}}].

\bibitem{Salmi:2012ta}
P.~Salmi and M.~Hindmarsh, \emph{{Radiation and Relaxation of Oscillons}},
  \href{https://doi.org/10.1103/PhysRevD.85.085033}{\emph{Phys. Rev. D}
  {\bfseries 85} (2012) 085033}
  [\href{https://arxiv.org/abs/1201.1934}{{\ttfamily 1201.1934}}].

\bibitem{Saffin:2014yka}
P.~M. Saffin, P.~Tognarelli and A.~Tranberg, \emph{{Oscillon Lifetime in the
  Presence of Quantum Fluctuations}},
  \href{https://doi.org/10.1007/JHEP08(2014)125}{\emph{JHEP} {\bfseries 08}
  (2014) 125} [\href{https://arxiv.org/abs/1401.6168}{{\ttfamily 1401.6168}}].

\bibitem{Mukaida:2014oza}
K.~Mukaida and M.~Takimoto, \emph{{Correspondence of I- and Q-balls as
  Non-relativistic Condensates}},
  \href{https://doi.org/10.1088/1475-7516/2014/08/051}{\emph{JCAP} {\bfseries
  08} (2014) 051} [\href{https://arxiv.org/abs/1405.3233}{{\ttfamily
  1405.3233}}].

\bibitem{Kawasaki:2015vga}
M.~Kawasaki, F.~Takahashi and N.~Takeda, \emph{{Adiabatic Invariance of
  Oscillons/I-balls}},
  \href{https://doi.org/10.1103/PhysRevD.92.105024}{\emph{Phys. Rev. D}
  {\bfseries 92} (2015) 105024}
  [\href{https://arxiv.org/abs/1508.01028}{{\ttfamily 1508.01028}}].

\bibitem{Fodor:2008es}
G.~Fodor, P.~Forgacs, Z.~Horvath and A.~Lukacs, \emph{{Small amplitude
  quasi-breathers and oscillons}},
  \href{https://doi.org/10.1103/PhysRevD.78.025003}{\emph{Phys. Rev. D}
  {\bfseries 78} (2008) 025003}
  [\href{https://arxiv.org/abs/0802.3525}{{\ttfamily 0802.3525}}].

\bibitem{Braaten:2016kzc}
E.~Braaten, A.~Mohapatra and H.~Zhang, \emph{{Nonrelativistic Effective Field
  Theory for Axions}},
  \href{https://doi.org/10.1103/PhysRevD.94.076004}{\emph{Phys. Rev. D}
  {\bfseries 94} (2016) 076004}
  [\href{https://arxiv.org/abs/1604.00669}{{\ttfamily 1604.00669}}].

\bibitem{Namjoo:2017nia}
M.~H. Namjoo, A.~H. Guth and D.~I. Kaiser, \emph{{Relativistic Corrections to
  Nonrelativistic Effective Field Theories}},
  \href{https://doi.org/10.1103/PhysRevD.98.016011}{\emph{Phys. Rev. D}
  {\bfseries 98} (2018) 016011}
  [\href{https://arxiv.org/abs/1712.00445}{{\ttfamily 1712.00445}}].

\bibitem{Braaten:2018lmj}
E.~Braaten, A.~Mohapatra and H.~Zhang, \emph{{Classical Nonrelativistic
  Effective Field Theories for a Real Scalar Field}},
  \href{https://doi.org/10.1103/PhysRevD.98.096012}{\emph{Phys. Rev. D}
  {\bfseries 98} (2018) 096012}
  [\href{https://arxiv.org/abs/1806.01898}{{\ttfamily 1806.01898}}].

\bibitem{Levkov:2022egq}
D.~G. Levkov, V.~E. Maslov, E.~Y. Nugaev and A.~G. Panin, \emph{{An Effective
  Field Theory for large oscillons}},
  \href{https://doi.org/10.1007/JHEP12(2022)079}{\emph{JHEP} {\bfseries 12}
  (2022) 079} [\href{https://arxiv.org/abs/2208.04334}{{\ttfamily
  2208.04334}}].

\bibitem{Vaquero:2018tib}
A.~Vaquero, J.~Redondo and J.~Stadler, \emph{{Early seeds of axion
  miniclusters}},
  \href{https://doi.org/10.1088/1475-7516/2019/04/012}{\emph{JCAP} {\bfseries
  04} (2019) 012} [\href{https://arxiv.org/abs/1809.09241}{{\ttfamily
  1809.09241}}].

\bibitem{Arvanitaki:2019rax}
A.~Arvanitaki, S.~Dimopoulos, M.~Galanis, L.~Lehner, J.~O. Thompson and
  K.~Van~Tilburg, \emph{{Large-misalignment mechanism for the formation of
  compact axion structures: Signatures from the QCD axion to fuzzy dark
  matter}}, \href{https://doi.org/10.1103/PhysRevD.101.083014}{\emph{Phys. Rev.
  D} {\bfseries 101} (2020) 083014}
  [\href{https://arxiv.org/abs/1909.11665}{{\ttfamily 1909.11665}}].

\bibitem{Copeland:2002ku}
E.~J. Copeland, S.~Pascoli and A.~Rajantie, \emph{{Dynamics of tachyonic
  preheating after hybrid inflation}},
  \href{https://doi.org/10.1103/PhysRevD.65.103517}{\emph{Phys. Rev. D}
  {\bfseries 65} (2002) 103517}
  [\href{https://arxiv.org/abs/hep-ph/0202031}{{\ttfamily hep-ph/0202031}}].

\bibitem{Inomata:2019ivs}
K.~Inomata, K.~Kohri, T.~Nakama and T.~Terada, \emph{{Enhancement of
  Gravitational Waves Induced by Scalar Perturbations due to a Sudden
  Transition from an Early Matter Era to the Radiation Era}},
  \href{https://doi.org/10.1103/PhysRevD.100.043532}{\emph{Phys. Rev. D}
  {\bfseries 100} (2019) 043532}
  [\href{https://arxiv.org/abs/1904.12879}{{\ttfamily 1904.12879}}].

\bibitem{Lozanov:2022yoy}
K.~D. Lozanov and V.~Takhistov, \emph{{Enhanced Gravitational Waves from
  Inflaton Oscillons}},  \href{https://arxiv.org/abs/2204.07152}{{\ttfamily
  2204.07152}}.

\bibitem{Adshead:2021kvl}
P.~Adshead and K.~D. Lozanov, \emph{{Self-gravitating Vector Dark Matter}},
  \href{https://doi.org/10.1103/PhysRevD.103.103501}{\emph{Phys. Rev. D}
  {\bfseries 103} (2021) 103501}
  [\href{https://arxiv.org/abs/2101.07265}{{\ttfamily 2101.07265}}].

\bibitem{Jain:2021pnk}
M.~Jain and M.~A. Amin, \emph{{Polarized solitons in higher-spin wave dark
  matter}}, \href{https://doi.org/10.1103/PhysRevD.105.056019}{\emph{Phys. Rev.
  D} {\bfseries 105} (2022) 056019}
  [\href{https://arxiv.org/abs/2109.04892}{{\ttfamily 2109.04892}}].

\bibitem{Zhang:2021xxa}
H.-Y. Zhang, M.~Jain and M.~A. Amin, \emph{{Polarized vector oscillons}},
  \href{https://doi.org/10.1103/PhysRevD.105.096037}{\emph{Phys. Rev. D}
  {\bfseries 105} (2022) 096037}
  [\href{https://arxiv.org/abs/2111.08700}{{\ttfamily 2111.08700}}].

\bibitem{Amin:2023imi}
M.~A. Amin, A.~J. Long and E.~D. Schiappacasse, \emph{{Photons from dark photon
  solitons via parametric resonance}},
  \href{https://arxiv.org/abs/2301.11470}{{\ttfamily 2301.11470}}.

\bibitem{Gorghetto:2022sue}
M.~Gorghetto, E.~Hardy, J.~March-Russell, N.~Song and S.~M. West, \emph{{Dark
  photon stars: formation and role as dark matter substructure}},
  \href{https://doi.org/10.1088/1475-7516/2022/08/018}{\emph{JCAP} {\bfseries
  08} (2022) 018} [\href{https://arxiv.org/abs/2203.10100}{{\ttfamily
  2203.10100}}].

\bibitem{Amin:2022pzv}
M.~A. Amin, M.~Jain, R.~Karur and P.~Mocz, \emph{{Small-scale structure in
  vector dark matter}},
  \href{https://doi.org/10.1088/1475-7516/2022/08/014}{\emph{JCAP} {\bfseries
  08} (2022) 014} [\href{https://arxiv.org/abs/2203.11935}{{\ttfamily
  2203.11935}}].

\bibitem{Jain:2023ojg}
M.~Jain, M.~A. Amin, J.~Thomas and W.~Wanichwecharungruang, \emph{{Kinetic
  relaxation and Bose-star formation in multicomponent dark matter- I}},
  \href{https://arxiv.org/abs/2304.01985}{{\ttfamily 2304.01985}}.

\bibitem{Eby:2015hyx}
J.~Eby, P.~Suranyi and L.~C.~R. Wijewardhana, \emph{{The Lifetime of Axion
  Stars}}, \href{https://doi.org/10.1142/S0217732316500905}{\emph{Mod. Phys.
  Lett. A} {\bfseries 31} (2016) 1650090}
  [\href{https://arxiv.org/abs/1512.01709}{{\ttfamily 1512.01709}}].

\bibitem{Eby:2017azn}
J.~Eby, M.~Ma, P.~Suranyi and L.~C.~R. Wijewardhana, \emph{{Decay of Ultralight
  Axion Condensates}},
  \href{https://doi.org/10.1007/JHEP01(2018)066}{\emph{JHEP} {\bfseries 01}
  (2018) 066} [\href{https://arxiv.org/abs/1705.05385}{{\ttfamily
  1705.05385}}].

\bibitem{Ibe:2019vyo}
M.~Ibe, M.~Kawasaki, W.~Nakano and E.~Sonomoto, \emph{{Decay of I-ball/Oscillon
  in Classical Field Theory}},
  \href{https://doi.org/10.1007/JHEP04(2019)030}{\emph{JHEP} {\bfseries 04}
  (2019) 030} [\href{https://arxiv.org/abs/1901.06130}{{\ttfamily
  1901.06130}}].

\bibitem{Ibe:2019lzv}
M.~Ibe, M.~Kawasaki, W.~Nakano and E.~Sonomoto, \emph{{Fragileness of Exact
  I-ball/Oscillon}},
  \href{https://doi.org/10.1103/PhysRevD.100.125021}{\emph{Phys. Rev. D}
  {\bfseries 100} (2019) 125021}
  [\href{https://arxiv.org/abs/1908.11103}{{\ttfamily 1908.11103}}].

\bibitem{Zhang:2020bec}
H.-Y. Zhang, M.~A. Amin, E.~J. Copeland, P.~M. Saffin and K.~D. Lozanov,
  \emph{{Classical Decay Rates of Oscillons}},
  \href{https://doi.org/10.1088/1475-7516/2020/07/055}{\emph{JCAP} {\bfseries
  07} (2020) 055} [\href{https://arxiv.org/abs/2004.01202}{{\ttfamily
  2004.01202}}].

\bibitem{Zhang:2020ntm}
H.-Y. Zhang, \emph{{Gravitational effects on oscillon lifetimes}},
  \href{https://doi.org/10.1088/1475-7516/2021/03/102}{\emph{JCAP} {\bfseries
  03} (2021) 102} [\href{https://arxiv.org/abs/2011.11720}{{\ttfamily
  2011.11720}}].

\bibitem{Grandclement:2011wz}
P.~Grandclement, G.~Fodor and P.~Forgacs, \emph{{Numerical simulation of
  oscillatons: extracting the radiating tail}},
  \href{https://doi.org/10.1103/PhysRevD.84.065037}{\emph{Phys. Rev. D}
  {\bfseries 84} (2011) 065037}
  [\href{https://arxiv.org/abs/1107.2791}{{\ttfamily 1107.2791}}].

\bibitem{Fodor:2009kg}
G.~Fodor, P.~Forgacs and M.~Mezei, \emph{{Mass loss and longevity of
  gravitationally bound oscillating scalar lumps (oscillatons) in
  D-dimensions}}, \href{https://doi.org/10.1103/PhysRevD.81.064029}{\emph{Phys.
  Rev. D} {\bfseries 81} (2010) 064029}
  [\href{https://arxiv.org/abs/0912.5351}{{\ttfamily 0912.5351}}].

\bibitem{Page:2003rd}
D.~N. Page, \emph{{Classical and quantum decay of oscillatons: Oscillating
  selfgravitating real scalar field solitons}},
  \href{https://doi.org/10.1103/PhysRevD.70.023002}{\emph{Phys. Rev. D}
  {\bfseries 70} (2004) 023002}
  [\href{https://arxiv.org/abs/gr-qc/0310006}{{\ttfamily gr-qc/0310006}}].

\bibitem{Eby:2020ply}
J.~Eby, L.~Street, P.~Suranyi and L.~C.~R. Wijewardhana, \emph{{Global view of
  axion stars with nearly Planck-scale decay constants}},
  \href{https://doi.org/10.1103/PhysRevD.103.063043}{\emph{Phys. Rev. D}
  {\bfseries 103} (2021) 063043}
  [\href{https://arxiv.org/abs/2011.09087}{{\ttfamily 2011.09087}}].

\bibitem{Hertzberg:2010yz}
M.~P. Hertzberg, \emph{{Quantum Radiation of Oscillons}},
  \href{https://doi.org/10.1103/PhysRevD.82.045022}{\emph{Phys. Rev. D}
  {\bfseries 82} (2010) 045022}
  [\href{https://arxiv.org/abs/1003.3459}{{\ttfamily 1003.3459}}].

\bibitem{Kawasaki:2013awa}
M.~Kawasaki and M.~Yamada, \emph{{Decay rates of Gaussian-type I-balls and
  Bose-enhancement effects in 3+1 dimensions}},
  \href{https://doi.org/10.1088/1475-7516/2014/02/001}{\emph{JCAP} {\bfseries
  02} (2014) 001} [\href{https://arxiv.org/abs/1311.0985}{{\ttfamily
  1311.0985}}].

\bibitem{Hertzberg:2018zte}
M.~P. Hertzberg and E.~D. Schiappacasse, \emph{{Dark Matter Axion Clump
  Resonance of Photons}},
  \href{https://doi.org/10.1088/1475-7516/2018/11/004}{\emph{JCAP} {\bfseries
  11} (2018) 004} [\href{https://arxiv.org/abs/1805.00430}{{\ttfamily
  1805.00430}}].

\bibitem{Ema:2015dka}
Y.~Ema, R.~Jinno, K.~Mukaida and K.~Nakayama, \emph{{Gravitational Effects on
  Inflaton Decay}},
  \href{https://doi.org/10.1088/1475-7516/2015/05/038}{\emph{JCAP} {\bfseries
  05} (2015) 038} [\href{https://arxiv.org/abs/1502.02475}{{\ttfamily
  1502.02475}}].

\bibitem{Ema:2016hlw}
Y.~Ema, R.~Jinno, K.~Mukaida and K.~Nakayama, \emph{{Gravitational particle
  production in oscillating backgrounds and its cosmological implications}},
  \href{https://doi.org/10.1103/PhysRevD.94.063517}{\emph{Phys. Rev. D}
  {\bfseries 94} (2016) 063517}
  [\href{https://arxiv.org/abs/1604.08898}{{\ttfamily 1604.08898}}].

\bibitem{Tang:2017hvq}
Y.~Tang and Y.-L. Wu, \emph{{On Thermal Gravitational Contribution to Particle
  Production and Dark Matter}},
  \href{https://doi.org/10.1016/j.physletb.2017.10.034}{\emph{Phys. Lett. B}
  {\bfseries 774} (2017) 676}
  [\href{https://arxiv.org/abs/1708.05138}{{\ttfamily 1708.05138}}].

\bibitem{Ema:2018ucl}
Y.~Ema, K.~Nakayama and Y.~Tang, \emph{{Production of Purely Gravitational Dark
  Matter}}, \href{https://doi.org/10.1007/JHEP09(2018)135}{\emph{JHEP}
  {\bfseries 09} (2018) 135}
  [\href{https://arxiv.org/abs/1804.07471}{{\ttfamily 1804.07471}}].

\bibitem{Chung:2018ayg}
D.~J.~H. Chung, E.~W. Kolb and A.~J. Long, \emph{{Gravitational production of
  super-Hubble-mass particles: an analytic approach}},
  \href{https://doi.org/10.1007/JHEP01(2019)189}{\emph{JHEP} {\bfseries 01}
  (2019) 189} [\href{https://arxiv.org/abs/1812.00211}{{\ttfamily
  1812.00211}}].

\bibitem{Mambrini:2021zpp}
Y.~Mambrini and K.~A. Olive, \emph{{Gravitational Production of Dark Matter
  during Reheating}},
  \href{https://doi.org/10.1103/PhysRevD.103.115009}{\emph{Phys. Rev. D}
  {\bfseries 103} (2021) 115009}
  [\href{https://arxiv.org/abs/2102.06214}{{\ttfamily 2102.06214}}].

\bibitem{Basso:2021whd}
E.~E. Basso and D.~J.~H. Chung, \emph{{Computation of gravitational particle
  production using adiabatic invariants}},
  \href{https://doi.org/10.1007/JHEP11(2021)146}{\emph{JHEP} {\bfseries 11}
  (2021) 146} [\href{https://arxiv.org/abs/2108.01653}{{\ttfamily
  2108.01653}}].

\bibitem{Clery:2021bwz}
S.~Clery, Y.~Mambrini, K.~A. Olive and S.~Verner, \emph{{Gravitational portals
  in the early Universe}},
  \href{https://doi.org/10.1103/PhysRevD.105.075005}{\emph{Phys. Rev. D}
  {\bfseries 105} (2022) 075005}
  [\href{https://arxiv.org/abs/2112.15214}{{\ttfamily 2112.15214}}].

\bibitem{Haque:2021mab}
M.~R. Haque and D.~Maity, \emph{{Gravitational dark matter: Free streaming and
  phase space distribution}},
  \href{https://doi.org/10.1103/PhysRevD.106.023506}{\emph{Phys. Rev. D}
  {\bfseries 106} (2022) 023506}
  [\href{https://arxiv.org/abs/2112.14668}{{\ttfamily 2112.14668}}].

\bibitem{Schiappacasse:2016nei}
E.~D. Schiappacasse and L.~H. Ford, \emph{{Graviton Creation by Small Scale
  Factor Oscillations in an Expanding Universe}},
  \href{https://doi.org/10.1103/PhysRevD.94.084030}{\emph{Phys. Rev. D}
  {\bfseries 94} (2016) 084030}
  [\href{https://arxiv.org/abs/1602.08416}{{\ttfamily 1602.08416}}].

\bibitem{Ema:2020ggo}
Y.~Ema, R.~Jinno and K.~Nakayama, \emph{{High-frequency Graviton from Inflaton
  Oscillation}},
  \href{https://doi.org/10.1088/1475-7516/2020/09/015}{\emph{JCAP} {\bfseries
  09} (2020) 015} [\href{https://arxiv.org/abs/2006.09972}{{\ttfamily
  2006.09972}}].

\bibitem{Garny:2015sjg}
M.~Garny, M.~Sandora and M.~S. Sloth, \emph{{Planckian Interacting Massive
  Particles as Dark Matter}},
  \href{https://doi.org/10.1103/PhysRevLett.116.101302}{\emph{Phys. Rev. Lett.}
  {\bfseries 116} (2016) 101302}
  [\href{https://arxiv.org/abs/1511.03278}{{\ttfamily 1511.03278}}].

\bibitem{Tang:2016vch}
Y.~Tang and Y.-L. Wu, \emph{{Pure Gravitational Dark Matter, Its Mass and
  Signatures}},
  \href{https://doi.org/10.1016/j.physletb.2016.05.045}{\emph{Phys. Lett. B}
  {\bfseries 758} (2016) 402}
  [\href{https://arxiv.org/abs/1604.04701}{{\ttfamily 1604.04701}}].

\bibitem{Garny:2017kha}
M.~Garny, A.~Palessandro, M.~Sandora and M.~S. Sloth, \emph{{Theory and
  Phenomenology of Planckian Interacting Massive Particles as Dark Matter}},
  \href{https://doi.org/10.1088/1475-7516/2018/02/027}{\emph{JCAP} {\bfseries
  02} (2018) 027} [\href{https://arxiv.org/abs/1709.09688}{{\ttfamily
  1709.09688}}].

\bibitem{Gross:2020zam}
C.~Gross, S.~Karamitsos, G.~Landini and A.~Strumia, \emph{{Gravitational Vector
  Dark Matter}}, \href{https://doi.org/10.1007/JHEP03(2021)174}{\emph{JHEP}
  {\bfseries 03} (2021) 174}
  [\href{https://arxiv.org/abs/2012.12087}{{\ttfamily 2012.12087}}].

\bibitem{Ford:1986sy}
L.~H. Ford, \emph{{Gravitational Particle Creation and Inflation}},
  \href{https://doi.org/10.1103/PhysRevD.35.2955}{\emph{Phys. Rev. D}
  {\bfseries 35} (1987) 2955}.

\bibitem{Chung:2001cb}
D.~J.~H. Chung, P.~Crotty, E.~W. Kolb and A.~Riotto, \emph{{On the
  Gravitational Production of Superheavy Dark Matter}},
  \href{https://doi.org/10.1103/PhysRevD.64.043503}{\emph{Phys. Rev. D}
  {\bfseries 64} (2001) 043503}
  [\href{https://arxiv.org/abs/hep-ph/0104100}{{\ttfamily hep-ph/0104100}}].

\bibitem{Chung:2011ck}
D.~J.~H. Chung, L.~L. Everett, H.~Yoo and P.~Zhou, \emph{{Gravitational Fermion
  Production in Inflationary Cosmology}},
  \href{https://doi.org/10.1016/j.physletb.2012.04.066}{\emph{Phys. Lett. B}
  {\bfseries 712} (2012) 147}
  [\href{https://arxiv.org/abs/1109.2524}{{\ttfamily 1109.2524}}].

\bibitem{Hashiba:2018tbu}
S.~Hashiba and J.~Yokoyama, \emph{{Gravitational particle creation for dark
  matter and reheating}},
  \href{https://doi.org/10.1103/PhysRevD.99.043008}{\emph{Phys. Rev. D}
  {\bfseries 99} (2019) 043008}
  [\href{https://arxiv.org/abs/1812.10032}{{\ttfamily 1812.10032}}].

\bibitem{Coleman:1985ki}
S.~R. Coleman, \emph{{Q-balls}},
  \href{https://doi.org/10.1016/0550-3213(86)90520-1}{\emph{Nucl. Phys. B}
  {\bfseries 262} (1985) 263}.

\bibitem{Friedberg:1986tp}
R.~Friedberg, T.~D. Lee and Y.~Pang, \emph{{MINI - SOLITON STARS}},
  \href{https://doi.org/10.1103/PhysRevD.35.3640}{\emph{Phys. Rev. D}
  {\bfseries 35} (1987) 3640}.

\bibitem{Cohen:1986ct}
A.~G. Cohen, S.~R. Coleman, H.~Georgi and A.~Manohar, \emph{{The Evaporation of
  $Q$ Balls}}, \href{https://doi.org/10.1016/0550-3213(86)90004-0}{\emph{Nucl.
  Phys. B} {\bfseries 272} (1986) 301}.

\bibitem{Kusenko:1997ad}
A.~Kusenko, \emph{{Small Q balls}},
  \href{https://doi.org/10.1016/S0370-2693(97)00582-0}{\emph{Phys. Lett. B}
  {\bfseries 404} (1997) 285}
  [\href{https://arxiv.org/abs/hep-th/9704073}{{\ttfamily hep-th/9704073}}].

\bibitem{Kusenko:1997zq}
A.~Kusenko, \emph{{Solitons in the supersymmetric extensions of the standard
  model}}, \href{https://doi.org/10.1016/S0370-2693(97)00584-4}{\emph{Phys.
  Lett. B} {\bfseries 405} (1997) 108}
  [\href{https://arxiv.org/abs/hep-ph/9704273}{{\ttfamily hep-ph/9704273}}].

\bibitem{Kusenko:1997si}
A.~Kusenko and M.~E. Shaposhnikov, \emph{{Supersymmetric Q balls as dark
  matter}}, \href{https://doi.org/10.1016/S0370-2693(97)01375-0}{\emph{Phys.
  Lett. B} {\bfseries 418} (1998) 46}
  [\href{https://arxiv.org/abs/hep-ph/9709492}{{\ttfamily hep-ph/9709492}}].

\bibitem{Affleck:1984fy}
I.~Affleck and M.~Dine, \emph{{A New Mechanism for Baryogenesis}},
  \href{https://doi.org/10.1016/0550-3213(85)90021-5}{\emph{Nucl. Phys. B}
  {\bfseries 249} (1985) 361}.

\bibitem{Dine:1995kz}
M.~Dine, L.~Randall and S.~D. Thomas, \emph{{Baryogenesis from flat directions
  of the supersymmetric standard model}},
  \href{https://doi.org/10.1016/0550-3213(95)00538-2}{\emph{Nucl. Phys. B}
  {\bfseries 458} (1996) 291}
  [\href{https://arxiv.org/abs/hep-ph/9507453}{{\ttfamily hep-ph/9507453}}].

\bibitem{Dolgov:1989us}
A.~D. Dolgov and D.~P. Kirilova, \emph{{ON PARTICLE CREATION BY A TIME
  DEPENDENT SCALAR FIELD}}, {\emph{Sov. J. Nucl. Phys.} {\bfseries 51} (1990)
  172}.

\bibitem{Traschen:1990sw}
J.~H. Traschen and R.~H. Brandenberger, \emph{{Particle Production During
  Out-of-equilibrium Phase Transitions}},
  \href{https://doi.org/10.1103/PhysRevD.42.2491}{\emph{Phys. Rev. D}
  {\bfseries 42} (1990) 2491}.

\bibitem{Shtanov:1994ce}
Y.~Shtanov, J.~H. Traschen and R.~H. Brandenberger, \emph{{Universe reheating
  after inflation}},
  \href{https://doi.org/10.1103/PhysRevD.51.5438}{\emph{Phys. Rev. D}
  {\bfseries 51} (1995) 5438}
  [\href{https://arxiv.org/abs/hep-ph/9407247}{{\ttfamily hep-ph/9407247}}].

\bibitem{Kofman:1997yn}
L.~Kofman, A.~D. Linde and A.~A. Starobinsky, \emph{{Towards the theory of
  reheating after inflation}},
  \href{https://doi.org/10.1103/PhysRevD.56.3258}{\emph{Phys. Rev. D}
  {\bfseries 56} (1997) 3258}
  [\href{https://arxiv.org/abs/hep-ph/9704452}{{\ttfamily hep-ph/9704452}}].

\bibitem{Kawasaki:2005xc}
M.~Kawasaki, K.~Konya and F.~Takahashi, \emph{{Q-ball instability due to U(1)
  breaking}}, \href{https://doi.org/10.1016/j.physletb.2005.05.082}{\emph{Phys.
  Lett. B} {\bfseries 619} (2005) 233}
  [\href{https://arxiv.org/abs/hep-ph/0504105}{{\ttfamily hep-ph/0504105}}].

\bibitem{Hiramatsu:2010dx}
T.~Hiramatsu, M.~Kawasaki and F.~Takahashi, \emph{{Numerical study of Q-ball
  formation in gravity mediation}},
  \href{https://doi.org/10.1088/1475-7516/2010/06/008}{\emph{JCAP} {\bfseries
  06} (2010) 008} [\href{https://arxiv.org/abs/1003.1779}{{\ttfamily
  1003.1779}}].

\bibitem{Hasegawa:2019bbo}
F.~Hasegawa, J.-P. Hong and M.~Suzuki, \emph{{More about Q-ball with elliptical
  orbit}}, \href{https://doi.org/10.1016/j.physletb.2019.135001}{\emph{Phys.
  Lett. B} {\bfseries 798} (2019) 135001}
  [\href{https://arxiv.org/abs/1903.07281}{{\ttfamily 1903.07281}}].

\bibitem{Co:2019wyp}
R.~T. Co and K.~Harigaya, \emph{{Axiogenesis}},
  \href{https://doi.org/10.1103/PhysRevLett.124.111602}{\emph{Phys. Rev. Lett.}
  {\bfseries 124} (2020) 111602}
  [\href{https://arxiv.org/abs/1910.02080}{{\ttfamily 1910.02080}}].

\bibitem{Co:2019jts}
R.~T. Co, L.~J. Hall and K.~Harigaya, \emph{{Axion Kinetic Misalignment
  Mechanism}},
  \href{https://doi.org/10.1103/PhysRevLett.124.251802}{\emph{Phys. Rev. Lett.}
  {\bfseries 124} (2020) 251802}
  [\href{https://arxiv.org/abs/1910.14152}{{\ttfamily 1910.14152}}].

\bibitem{Zhou:2013tsa}
S.-Y. Zhou, E.~J. Copeland, R.~Easther, H.~Finkel, Z.-G. Mou and P.~M. Saffin,
  \emph{{Gravitational Waves from Oscillon Preheating}},
  \href{https://doi.org/10.1007/JHEP10(2013)026}{\emph{JHEP} {\bfseries 10}
  (2013) 026} [\href{https://arxiv.org/abs/1304.6094}{{\ttfamily 1304.6094}}].

\bibitem{Antusch:2016con}
S.~Antusch, F.~Cefala and S.~Orani, \emph{{Gravitational waves from oscillons
  after inflation}},
  \href{https://doi.org/10.1103/PhysRevLett.118.011303}{\emph{Phys. Rev. Lett.}
  {\bfseries 118} (2017) 011303}
  [\href{https://arxiv.org/abs/1607.01314}{{\ttfamily 1607.01314}}].

\bibitem{Liu:2017hua}
J.~Liu, Z.-K. Guo, R.-G. Cai and G.~Shiu, \emph{{Gravitational Waves from
  Oscillons with Cuspy Potentials}},
  \href{https://doi.org/10.1103/PhysRevLett.120.031301}{\emph{Phys. Rev. Lett.}
  {\bfseries 120} (2018) 031301}
  [\href{https://arxiv.org/abs/1707.09841}{{\ttfamily 1707.09841}}].

\bibitem{Lozanov:2017hjm}
K.~D. Lozanov and M.~A. Amin, \emph{{Self-resonance after inflation: oscillons,
  transients and radiation domination}},
  \href{https://doi.org/10.1103/PhysRevD.97.023533}{\emph{Phys. Rev. D}
  {\bfseries 97} (2018) 023533}
  [\href{https://arxiv.org/abs/1710.06851}{{\ttfamily 1710.06851}}].

\bibitem{Amin:2018xfe}
M.~A. Amin, J.~Braden, E.~J. Copeland, J.~T. Giblin, C.~Solorio, Z.~J. Weiner
  et~al., \emph{{Gravitational waves from asymmetric oscillon dynamics?}},
  \href{https://doi.org/10.1103/PhysRevD.98.024040}{\emph{Phys. Rev. D}
  {\bfseries 98} (2018) 024040}
  [\href{https://arxiv.org/abs/1803.08047}{{\ttfamily 1803.08047}}].

\bibitem{Kitajima:2018zco}
N.~Kitajima, J.~Soda and Y.~Urakawa, \emph{{Gravitational wave forest from
  string axiverse}},
  \href{https://doi.org/10.1088/1475-7516/2018/10/008}{\emph{JCAP} {\bfseries
  10} (2018) 008} [\href{https://arxiv.org/abs/1807.07037}{{\ttfamily
  1807.07037}}].

\bibitem{Liu:2018rrt}
J.~Liu, Z.-K. Guo, R.-G. Cai and G.~Shiu, \emph{{Gravitational wave production
  after inflation with cuspy potentials}},
  \href{https://doi.org/10.1103/PhysRevD.99.103506}{\emph{Phys. Rev. D}
  {\bfseries 99} (2019) 103506}
  [\href{https://arxiv.org/abs/1812.09235}{{\ttfamily 1812.09235}}].

\bibitem{Lozanov:2019ylm}
K.~D. Lozanov and M.~A. Amin, \emph{{Gravitational perturbations from oscillons
  and transients after inflation}},
  \href{https://doi.org/10.1103/PhysRevD.99.123504}{\emph{Phys. Rev. D}
  {\bfseries 99} (2019) 123504}
  [\href{https://arxiv.org/abs/1902.06736}{{\ttfamily 1902.06736}}].

\bibitem{Hiramatsu:2020obh}
T.~Hiramatsu, E.~I. Sfakianakis and M.~Yamaguchi, \emph{{Gravitational wave
  spectra from oscillon formation after inflation}},
  \href{https://doi.org/10.1007/JHEP03(2021)021}{\emph{JHEP} {\bfseries 03}
  (2021) 021} [\href{https://arxiv.org/abs/2011.12201}{{\ttfamily
  2011.12201}}].

\bibitem{Kou:2021bij}
X.-X. Kou, J.~B. Mertens, C.~Tian and S.-Y. Zhou, \emph{{Gravitational waves
  from fully general relativistic oscillon preheating}},
  \href{https://doi.org/10.1103/PhysRevD.105.123505}{\emph{Phys. Rev. D}
  {\bfseries 105} (2022) 123505}
  [\href{https://arxiv.org/abs/2112.07626}{{\ttfamily 2112.07626}}].

\bibitem{Garcia:2023eol}
M.~A.~G. Garcia and M.~Pierre, \emph{{Reheating after Inflaton Fragmentation}},
   \href{https://arxiv.org/abs/2306.08038}{{\ttfamily 2306.08038}}.

\bibitem{Kusenko:2008zm}
A.~Kusenko and A.~Mazumdar, \emph{{Gravitational waves from fragmentation of a
  primordial scalar condensate into Q-balls}},
  \href{https://doi.org/10.1103/PhysRevLett.101.211301}{\emph{Phys. Rev. Lett.}
  {\bfseries 101} (2008) 211301}
  [\href{https://arxiv.org/abs/0807.4554}{{\ttfamily 0807.4554}}].

\bibitem{Kusenko:2009cv}
A.~Kusenko, A.~Mazumdar and T.~Multamaki, \emph{{Gravitational waves from the
  fragmentation of a supersymmetric condensate}},
  \href{https://doi.org/10.1103/PhysRevD.79.124034}{\emph{Phys. Rev. D}
  {\bfseries 79} (2009) 124034}
  [\href{https://arxiv.org/abs/0902.2197}{{\ttfamily 0902.2197}}].

\bibitem{Chiba:2009zu}
T.~Chiba, K.~Kamada and M.~Yamaguchi, \emph{{Gravitational Waves from Q-ball
  Formation}}, \href{https://doi.org/10.1103/PhysRevD.81.083503}{\emph{Phys.
  Rev. D} {\bfseries 81} (2010) 083503}
  [\href{https://arxiv.org/abs/0912.3585}{{\ttfamily 0912.3585}}].

\bibitem{Ema:2021fdz}
Y.~Ema, K.~Mukaida and K.~Nakayama, \emph{{Scalar field couplings to quadratic
  curvature and decay into gravitons}},
  \href{https://doi.org/10.1007/JHEP05(2022)087}{\emph{JHEP} {\bfseries 05}
  (2022) 087} [\href{https://arxiv.org/abs/2112.12774}{{\ttfamily
  2112.12774}}].

\bibitem{Ejlli:2019bqj}
A.~Ejlli, D.~Ejlli, A.~M. Cruise, G.~Pisano and H.~Grote, \emph{{Upper limits
  on the amplitude of ultra-high-frequency gravitational waves from graviton to
  photon conversion}},
  \href{https://doi.org/10.1140/epjc/s10052-019-7542-5}{\emph{Eur. Phys. J. C}
  {\bfseries 79} (2019) 1032}
  [\href{https://arxiv.org/abs/1908.00232}{{\ttfamily 1908.00232}}].

\bibitem{Ringwald:2020ist}
A.~Ringwald, J.~Schutte-Engel and C.~Tamarit, \emph{{Gravitational Waves as a
  Big Bang Thermometer}},
  \href{https://doi.org/10.1088/1475-7516/2021/03/054}{\emph{JCAP} {\bfseries
  03} (2021) 054} [\href{https://arxiv.org/abs/2011.04731}{{\ttfamily
  2011.04731}}].

\bibitem{Aggarwal:2020olq}
N.~Aggarwal et~al., \emph{{Challenges and opportunities of gravitational-wave
  searches at MHz to GHz frequencies}},
  \href{https://doi.org/10.1007/s41114-021-00032-5}{\emph{Living Rev. Rel.}
  {\bfseries 24} (2021) 4} [\href{https://arxiv.org/abs/2011.12414}{{\ttfamily
  2011.12414}}].

\bibitem{Berlin:2021txa}
A.~Berlin, D.~Blas, R.~Tito~D'Agnolo, S.~A.~R. Ellis, R.~Harnik, Y.~Kahn
  et~al., \emph{{Detecting high-frequency gravitational waves with microwave
  cavities}}, \href{https://doi.org/10.1103/PhysRevD.105.116011}{\emph{Phys.
  Rev. D} {\bfseries 105} (2022) 116011}
  [\href{https://arxiv.org/abs/2112.11465}{{\ttfamily 2112.11465}}].

\bibitem{Domcke:2022rgu}
V.~Domcke, C.~Garcia-Cely and N.~L. Rodd, \emph{{Novel Search for
  High-Frequency Gravitational Waves with Low-Mass Axion Haloscopes}},
  \href{https://doi.org/10.1103/PhysRevLett.129.041101}{\emph{Phys. Rev. Lett.}
  {\bfseries 129} (2022) 041101}
  [\href{https://arxiv.org/abs/2202.00695}{{\ttfamily 2202.00695}}].

\bibitem{Tobar:2022pie}
M.~E. Tobar, C.~A. Thomson, W.~M. Campbell, A.~Quiskamp, J.~F. Bourhill, B.~T.
  McAllister et~al., \emph{{Comparing Instrument Spectral Sensitivity of
  Dissimilar Electromagnetic Haloscopes to Axion Dark Matter and High Frequency
  Gravitational Waves}},
  \href{https://doi.org/10.3390/sym14102165}{\emph{Symmetry} {\bfseries 14}
  (2022) 2165} [\href{https://arxiv.org/abs/2209.03004}{{\ttfamily
  2209.03004}}].

\bibitem{Dolgov:2012be}
A.~D. Dolgov and D.~Ejlli, \emph{{Conversion of relic gravitational waves into
  photons in cosmological magnetic fields}},
  \href{https://doi.org/10.1088/1475-7516/2012/12/003}{\emph{JCAP} {\bfseries
  12} (2012) 003} [\href{https://arxiv.org/abs/1211.0500}{{\ttfamily
  1211.0500}}].

\bibitem{Domcke:2020yzq}
V.~Domcke and C.~Garcia-Cely, \emph{{Potential of radio telescopes as
  high-frequency gravitational wave detectors}},
  \href{https://doi.org/10.1103/PhysRevLett.126.021104}{\emph{Phys. Rev. Lett.}
  {\bfseries 126} (2021) 021104}
  [\href{https://arxiv.org/abs/2006.01161}{{\ttfamily 2006.01161}}].

\bibitem{Ramazanov:2023nxz}
S.~Ramazanov, R.~Samanta, G.~Trenkler and F.~R. Urban, \emph{{Shimmering
  gravitons in the gamma-ray sky}},
  \href{https://arxiv.org/abs/2304.11222}{{\ttfamily 2304.11222}}.

\bibitem{Liu:2023mll}
T.~Liu, J.~Ren and C.~Zhang, \emph{{Detecting High-Frequency Gravitational
  Waves in Planetary Magnetosphere}},
  \href{https://arxiv.org/abs/2305.01832}{{\ttfamily 2305.01832}}.

\bibitem{Ito:2023fcr}
A.~Ito, K.~Kohri and K.~Nakayama, \emph{{Probing high frequency gravitational
  waves with pulsars}},  \href{https://arxiv.org/abs/2305.13984}{{\ttfamily
  2305.13984}}.

\end{thebibliography}\endgroup

\end{document}